\documentclass[12pt]{article}
\usepackage{amsmath}
\usepackage{cite}
\usepackage[dvipdfmx]{graphicx}
\usepackage{color}
\begin{document}
\begin{flushright}
KANAZAWA-21-06\\
May, 2021
\end{flushright}
\vspace*{1cm}

\renewcommand\thefootnote{\fnsymbol{footnote}}
\begin{center} 
 {\Large\bf Inflation connected to the origin of $CP$ violation}
\vspace*{1cm}

{\large Tsuyoshi Hashimoto}$^a$\footnote[1]{e-mail:~ 
t\_hashimoto@hep.s.kanazawa-u.ac.jp}, 
{\large Norma Sidik Risdianto}$^{a,b}$\footnote[2]{e-mail:~ 
norma.risdianto@uin-suka.ac.id}\\
{\Large and Daijiro Suematsu}$^a$\footnote[3]{e-mail:
~suematsu@hep.s.kanazawa-u.ac.jp}
\vspace*{0.5cm}\\

$^a${\it Institute for Theoretical Physics, Kanazawa University, 
Kanazawa 920-1192, Japan}

$^b${\it Department of Physics Education, Universitas Islam 
Negeri Sunan Kalijaga, Jl. 
Marsda Adisucipto, 55280, Yogyakarta, Indonesia}
\end{center}
\vspace*{1.5cm} 

\noindent
{\Large\bf Abstract}\\
We consider a simple extension of the standard model, which could give a solution for
its $CP$ issues such as the origin of  both CKM and PMNS phases and the strong $CP$ 
problem. The model is extended with singlet scalars which 
allow the introduction of the Peccei-Quinn symmetry and
could cause spontaneous $CP$ violation to result in these phases at low energy regions. 
The singlet scalars 
could give a good inflaton candidate if they have a suitable nonminimal coupling with 
the Ricci scalar. $CP$ issues and inflation could be closely related through these singlet 
scalars in a natural way. In a case where inflaton is a mixture of the singlet scalars,
we study reheating and leptogenesis as notable phenomena affected by the fields 
introduced in this extension. 
 
\newpage
\setcounter{footnote}{0}
\renewcommand\thefootnote{\alph{footnote}}

\section{Introduction}
$CP$ symmetry is a fundamental discrete symmetry which plays an important role 
in particle physics. In the standard model (SM), it is considered to be violated 
explicitly through complex Yukawa coupling constants \cite{km} and a $\theta$ 
parameter in the QCD sector \cite{theta}. The former is known to explain very well 
$CP$ violating phenomena in $B$ meson systems and so on \cite{pdg}. 
The latter is severely constrained through an experimental search of 
a neutron electric dipole moment \cite{ediploe} and causes the notorious 
strong $CP$ problem \cite{strongcp}. 
Peccei-Quinn (PQ) symmetry has been proposed to solve it \cite{pq}.
If we assume that the $CP$ symmetry is an original symmetry of nature, 
the complex phases of Yukawa coupling constants have to be spontaneously 
induced through some mechanism at high energy regions.
It may be compactification dynamics in string theory near at the Planck 
scale \cite{comp-cp}. 
In that case, since nonzero $\theta$ could be caused through radiative effects after 
the $CP$ violation, the PQ symmetry is required to solve the strong $CP$ 
problem again. 

As an alternative scenario for the realization of $CP$ symmetry, 
we may consider it to be exact so that all coupling constants 
including Yukawa couplings are real and also $\theta=0$ is kept 
to some scale much lower than the Planck scale.
In that case, the $CP$ symmetry is supposed to be spontaneously broken and 
this violation can be expected to be transformed to a complex phase in the CKM 
matrix effectively. 
If nonzero $\theta$ is not brought about in this process, it is favorable for 
the strong $CP$ problem. The Nelson-Barr (NB) mechanism \cite{nb} has 
been proposed as such a concrete example. 
Unfortunately, radiative effects could cause nonzero $\theta$ with a magnitude 
which contradicts the experimental constraints \cite{nb-theta}.\footnote{
Introduction of the PQ symmetry could solve this fault of the model.
We consider such a possibility in the extension of the model.} 
However, the scenario is interesting since it can present an explanation for
the origin of the $CP$ violation at a much lower energy region than the
Planck scale. As a realization of the NB mechanism, a simple model has been 
proposed in \cite{bbp}. The model is extended in \cite{patisalam} to the 
lepton sector where the existence of a $CP$ violating phase in the 
PMNS matrix \cite{pmns} is suggested through recent neutrino oscillation 
experiments \cite{lepcp}. 

Observations of the CMB fluctuation \cite{cmb,planck18} suggest the 
existence of the exponential expansion of the universe called inflation.
Inflation is usually considered to be induced by some slowly rolling scalar field 
called inflaton \cite{inf}. 
It is a crucial problem to identify its candidate from a viewpoint of the extension 
of the SM. Although the Higgs scalar has been studied as an only promising 
candidate in the SM \cite{higgsinf} under an assumption that 
it has a nonminimal coupling with the Ricci scalar curvature \cite{nonminimal}, 
several problems have been pointed out \cite{stab,unitarity,spike}.  
In this situation, it is interesting to find an alternative candidate for inflaton   
in a certain extension of the SM which could solve several problems in the SM. 
In this sense, the model extended from a viewpoint of the $CP$ 
symmetry as described above could give such a promising candidate.
It contains singlet scalars which cause the spontaneous $CP$ violation 
and allow the introduction of the PQ symmetry as a solution 
for the strong $CP$ problem. 
If they couple with the scalar curvature nonminimally, it could cause
slow-roll inflation successfully. 
In this paper, we discuss such a possibility that the inflation of the universe 
could be related to the $CP$ violation in the SM.
We study reheating and leptogenesis as its phenomenology caused 
by extra fields introduced in the model to solve the $CP$ issues.

Remaining parts of the paper are organized as follows.
In section 2, we describe the model studied in this paper and discuss 
both phases in the CKM and PMNS matrices which are derived as a result 
of the spontaneous $CP$ violation. In section 3, we discuss the inflation brought 
about by the singlet scalars which are related to the $CP$ issues and
the reheating. After that, we describe leptogenesis 
which could show a distinguishable feature from the usual leptogenesis in the
seesaw model.  The paper is summarized in section 4.

\section{Origin of $CP$ violation}
\subsection{An extended model}
Our model is an $CP$ invariant extension of the SM with global $U(1)\times Z_4$ 
symmetry and several additional fields.
As fermions, we introduce
 a pair of vector-like down-type quarks $(D_L, D_R)$, a pair of vector-like charged 
leptons $(E_L, E_R)$, and three right-handed singlet fermions 
$N_j~(j=1,2,3)$.\footnote{Similar models with vector-like extra fermions have 
been considered under different symmetry structures \cite{patisalam,mksvz}. } 
We also introduce an additional doublet scalar $\eta$ and two singlet complex 
scalars $\sigma$ and $S$. Their representation and charge under 
the symmetry $[SU(3)_C\times SU(2)_L\times U(1)_Y]\times U(1)\times Z_4$
are summarized in Table 1.\footnote{
$Z_4$ is imposed by hand to control the couplings of the new fields 
to the SM contents.} 
The SM contents are assumed to have no charge of the global symmetry.
Since this global $U(1)$ has color anomaly in the same way as the KSVZ 
model \cite{ksvz} for a strong $CP$ problem, 
it can play the role of the PQ symmetry.
The present charge assignment for colored fermions guarantees the domain wall 
number to be one ($N_{\rm DW}=1$) 
so that the model can escape the domain wall problem \cite{dw,stdw}.

\begin{figure}[t]
\begin{center}
\small
\begin{tabular}{c|ccccc||c|ccccc}\hline
& $SU(3)_C$ & $SU(2)_L$ & $U(1)_Y$ & $U(1)$ & $Z_4$ & &
$SU(3)_C$ & $SU(2)_L$ & $U(1)_Y$ & $U(1)$ & $Z_4$ \\\hline
$D_L$ & {\bf 3} & {\bf 1} & $-\frac{1}{3}$ & 0 & 2 & 
$D_R$ & {\bf 3} & {\bf 1} & $-\frac{1}{3}$ & 2 & 0 \\
$E_L$ & {\bf 1} & {\bf 1} & $-1$ & 0 & 2 & 
$E_R$ & {\bf 1} & {\bf 1} & $-1$ & 2 & 0 \\
$\sigma$ & {\bf 1} & {\bf 1} & 0 & $-2$ & 2 & 
$S$ & {\bf 1} & {\bf 1} & $0$ & 0 & 2 \\
$N_k$ & {\bf 1} & {\bf 1} & 0 & $1$ & 1 & 
$\eta$ & {\bf 1} & {\bf 2} & $-\frac{1}{2}$ & $-1$ &$-1$ \\\hline
\end{tabular}
\end{center}
{\footnotesize {\bf Table 1}~~New fields added to the SM and their 
representation and charge under 
$[SU(3)_C\times SU(2)_L\times U(1)_Y]\times U(1)\times Z_4$.} 
\end{figure}
\normalsize

The model is characterized by new Yukawa terms and
scalar potential which are invariant under the imposed symmetry
\begin{eqnarray}
-{\cal L}_Y&=&y_D\sigma\bar D_LD_R +y_E\sigma\bar E_LE_R+
\sum_{j=1}^3\Big(\frac{y_{N_j}}{2}\sigma \bar N_j^cN_j+ y_{d_j}S\bar D_L d_{R_j}
+ \tilde y_{d_j}S^\dagger\bar D_L d_{R_j} \nonumber\\ 
&+&y_{e_j}S\bar E_L e_{R_j} +\tilde y_{e_j}S^\dagger\bar E_L e_{R_j} +
\sum_{\alpha=1}^3 h_{\alpha j}^\ast\eta \bar\ell_\alpha N_j \Big)
+\sum_{\alpha,\beta=1}^3
\frac{y_{\alpha\beta}}{M_\ast}(\bar\ell_\alpha\phi)(\bar\ell_\beta\phi)
+ {\rm h.c.}, \nonumber \\
V&=&\lambda_1(\phi^\dagger\phi)^2+\lambda_2(\eta^\dagger\eta)^2
+\lambda_3(\phi^\dagger\phi)(\eta^\dagger\eta)+
\lambda_4(\phi^\dagger\eta)(\eta^\dagger\phi)+
\frac{\lambda_5}{2M_\ast}[\sigma(\eta^\dagger\phi)^2+{\rm h.c.}] \nonumber \\
&+&\kappa_\sigma(\sigma^\dagger\sigma)^2+\kappa_S(S^\dagger S)^2
+(\kappa_{\phi\sigma}\phi^\dagger\phi
+\kappa_{\eta\sigma}\eta^\dagger\eta)(\sigma^\dagger\sigma)
+(\kappa_{\phi S}\phi^\dagger\phi+\kappa_{\eta S}\eta^\dagger\eta) (S^\dagger S)
\nonumber \\
&+&\kappa_{\sigma S}(\sigma^\dagger\sigma)(S^\dagger S) 
+ m_\phi^2\phi^\dagger\phi+ m_\eta^2\eta^\dagger\eta
+m_\sigma^2\sigma^\dagger\sigma+m_S^2S^\dagger S +V_b,
\label{model}
\end{eqnarray} 
where $d_{R_j}$ and $e_{R_j}$ are the SM down-type quarks and charged leptons, respectively.
$\ell_\alpha$ is a doublet lepton and $\phi$ is an ordinary doublet Higgs scalar.
Since $CP$ invariance is assumed, parameters in Lagrangian are 
considered to be all real. 
In eq.~(\ref{model}), we list dominant terms up to dimension five and 
$M_\ast$ is a cut-off scale of the model. 
Other invariant terms are higher order and can be safely 
neglected in comparison with the listed ones. 
$V_b$ is composed of terms which are invariant under the global symmetry but 
violate the $S$ number.

For a while, we focus on a part of field space where the field values of $\sigma$
and $S$ are much larger than both $\phi$ and $\eta$ to study the potential
composed of $\sigma$ and $S$ only.
In the present study, we assume that $V_b$ takes a following form:\footnote{
Imposed symmetry allows terms 
$m_S^{\prime 2}(S^2+S^{\dagger 2})$ in $V_b$. However, we assume that their contribution 
is negligible since we focus our attention on a potential valley where 
$\tilde\sigma\propto \tilde S$ is satisfied and then $\cos 2\rho$ could be a constant 
at the minimum of $V_b$ in that case.} 
\begin{equation}
V_b=\alpha(S^4+S^{\dagger 4})+\beta\sigma^\dagger\sigma(S^2+S^{\dagger 2})
=\frac{1}{2}\tilde S^2(\alpha\tilde S^2\cos 4\rho+\beta\tilde\sigma^2\cos 2\rho),
\label{vb}
\end{equation}
where we define $\sigma=\frac{\tilde\sigma}{\sqrt 2} e^{i\theta}$ and 
$S=\frac{\tilde S}{\sqrt 2}e^{i\rho}$.
Along the minimum of $V_b$ for $\rho$ which is fixed by $\frac{\partial V_b}{\partial\rho}=0$, 
the potential of $\tilde\sigma$ and $\tilde S$ can be written as
\begin{equation}
V(\tilde\sigma, \tilde S)=\frac{\tilde\kappa_\sigma}{4}(\tilde \sigma^2-w^2)^2+
\frac{\tilde\kappa_S}{4}(\tilde S^2-u^2)^2+
\frac{\kappa_{\sigma S}}{4}(\tilde\sigma^2-w^2)(\tilde S^2-u^2),
\label{potv}
\end{equation} 
where $\tilde\kappa_\sigma$ and $\tilde \kappa_S$ are defined as
\begin{equation}
\tilde\kappa_\sigma=\kappa_\sigma-\frac{\beta^2}{4\alpha}, \qquad
\tilde\kappa_S=\kappa_S-2\alpha,
\end{equation}
and $w$ and $u$ are the vacuum expectation values (VEVs) of 
$\tilde\sigma$ and $\tilde S$.
They are supposed to be much larger than the weak scale.
They keep the gauge symmetry but break down the global symmetry $U(1)\times Z_4$ 
into its diagonal subgroup $Z_2$.\footnote{It guarantees the stability of the 
lightest $Z_2$ odd field, which could be a dark matter (DM) candidate as discussed later.}
Since the minimum of $V_b$ can be determined by using these VEVs as 
$\cos 2\rho=-\frac{\beta}{4\alpha}\frac{w^2}{u^2}$ 
as long as $\left|\frac{\beta}{4\alpha}\frac{w^2}{u^2}\right|\le1$ is satisfied, 
the $CP$ symmetry is spontaneously broken to result in a low energy effective 
model with the $CP$ violation. 
On the other hand, $\theta=0$ is satisfied because of 
the global $U(1)$ symmetry relevant to $\sigma$ \cite{ssbcp}. 
The stability condition for the potential (\ref{potv}) can be given as
\begin{equation}
\tilde\kappa_\sigma, \tilde\kappa_S>0, \qquad  
4\tilde\kappa_\sigma\tilde\kappa_S>\kappa_{\sigma S}^2.
\label{stab1}
\end{equation}

If we consider the fluctuation of $\tilde\sigma$ and $\tilde S$ around the vacua
$\langle\tilde\sigma\rangle$ and $\langle\tilde S\rangle$,
mass eigenstates are the mixture of them in general. 
If we take account of the stability condition (\ref{stab1}), 
mass eigenvalues can be approximately expressed as
\begin{eqnarray} 
&&m^2_{\tilde S}\simeq 2\left(\tilde\kappa_S-\frac{\kappa_{\sigma S}^2}
{4\tilde\kappa_\sigma}\right)u^2 \equiv 2\hat\kappa_Su^2, \quad 
m^2_{\tilde\sigma}\simeq 2\tilde\kappa_\sigma w^2 
\qquad {\rm for}~~\tilde \kappa_\sigma^2 w^2 \gg \tilde\kappa_S^2u^2,  \nonumber \\ 
&&m_{\tilde S}^2\simeq 2\tilde\kappa_S u^2, \quad m^2_{\tilde\sigma}
\simeq 2 \left(\tilde\kappa_\sigma-\frac{\kappa_{\sigma S}^2}
{4\tilde\kappa_S}\right)w^2\equiv 2\hat\kappa_\sigma w^2
\qquad {\rm for}~~\tilde \kappa_S^2 u^2 \gg \tilde\kappa_\sigma^2 w^2.
\end{eqnarray} 
Although they have a tiny subcomponent in these cases,
 a dominant component of their eigenstates 
is $\tilde S$ and $\tilde\sigma$, respectively. 
The mass of an orthogonal component to $\tilde S$ is found to 
be $m^2_{S_\perp}=8\alpha u^2\left(1-\cos^22\rho\right)$.
Since the global $U(1)$ symmetry works as the PQ symmetry mentioned above,
and the axion decay constant is given as $f_a=w$, 
the VEV $w$ should satisfy the following condition \cite{pdg,fa}: 
\begin{equation}
4\times 10^8~{\rm GeV}~{^<_\sim}~ w ~{^<_\sim}~10^{11} {\rm GeV}.
\end{equation}
The NG-boson caused by the spontaneous breaking of this $U(1)$ becomes axion 
\cite{axion} which is characterized by a coupling with photons \cite{lmn}
\begin{equation}
g_{a\gamma\gamma}=\frac{1.51}{10^{10}~{\rm GeV}}\left(\frac{m_a}{\rm eV}\right).
\end{equation} 

In the next part, we show that the effective model after the symmetry 
breaking can have 
 $CP$ phases in the CKM  and PMNS matrices. 
They are induced by the mass matrices for
 the down type quarks and the charged leptons  
through a similar mechanism which has been discussed in \cite{bbp}
as a simple realization of the NB mechanism \cite{nb} for the 
strong $CP$ problem.
 
\subsection{$CP$ violating phases in CKM and PMNS matrices}
Yukawa couplings of down-type quarks and charged leptons given in eq.~(\ref{model}) 
derive mass terms as 
\begin{equation}
(\bar f_{Li}, \bar F_L){\cal M}_f
\left(\begin{array}{c} f_{R_j} \\ F_R \\ \end{array} \right)+{\rm h.c.}, \qquad
{\cal M}_f=\left(\begin{array}{cc}
m_{f_{ij}} & 0 \\ {\cal F}_{f_j} & \mu_F \\
\end{array}\right), 
\label{dmass}
\end{equation}
where $f$ and $F$ represent $f=d,e$ and $F=D,E$ for down-type quarks and 
charged leptons and ${\cal M}_f$ is a $4\times 4$ matrix.
 Each component of ${\cal M}_f$ is expressed as
 $m_{f_{ij}}=h_{f_{ij}}\langle\tilde\phi\rangle$, 
${\cal F}_{f_j}=(y_{f_j} ue^{i\rho} + \tilde y_{f_j} 
ue^{-i\rho})$ and $\mu_F=y_F w$.  
This mass matrix is found to have the same form proposed in \cite{bbp}.
Since the global $U(1)$ symmetry works as the PQ symmetry and all parameters in the model 
are assumed to be real, ${\rm arg}({\rm det}{\cal M}_f)=0$ is satisfied even if 
radiative effects are taken into account after the spontaneous breaking of 
the $CP$ symmetry \cite{patisalam}.

We consider the diagonalization of a matrix ${\cal M}_f{\cal M}_f^\dagger$ 
by a unitary matrix
\begin{equation}
\left(\begin{array}{cc} A_f & B_f \\ C_f& D_f \\\end{array}\right)
\left(\begin{array}{cc} m_fm_f^{\dagger} & m_f{\cal F}_f^\dagger \\ 
  {\cal F}_fm_f^\dagger & \mu_F^2 +{\cal F}_f{\cal F}_f^\dagger \\
\end{array}\right)
\left(\begin{array}{cc} A_f^\dagger & C_f^\dagger \\ B_f^\dagger 
& D_f^\dagger \\\end{array}\right)=
\left(\begin{array}{cc} \tilde m^2_f & 0 \\ 0 & \tilde M^2_F \\\end{array}\right),
\label{mass}
\end{equation}
where $\tilde m^2_f$ is  a $3\times 3$ diagonal matrix in which generation 
indices are not explicitly written. Eq.~(\ref{mass}) requires
\begin{eqnarray}
   && m_fm_f^\dagger=A_f^\dagger \tilde m^2_f A_f+ 
C_f^\dagger \tilde M^2_FC_f, \qquad
  {\cal F}_fm_f^\dagger=B_f^\dagger m^2A_f+ D_f^\dagger \tilde M^2_FC_f, \nonumber \\
   && \mu_F^2+{\cal F}_f{\cal F}_f^\dagger=
B_f^\dagger \tilde m^2_f B_f+ D_f^\dagger \tilde M^2_F D_f.
\end{eqnarray}  
If $\mu_F^2+{\cal F}_f{\cal F}_f^\dagger$ is much larger than each 
component of ${\cal F}_fm_f^\dagger$, which can be realized in the case
$u, w\gg \langle \phi\rangle$,
we find that $B_f, C_f$, and $D_f$ are approximately given as
\begin{equation}
  B_f\simeq -\frac{A_fm_f{\cal F}_f^\dagger}{\mu_F^2
+{\cal F}_f{\cal F}_f^\dagger},
 \qquad C_f\simeq\frac{{\cal F}_f m_f^\dagger}{\mu_F^2
+{\cal F}_f{\cal F}_f^\dagger},
   \qquad D_f\simeq 1.
\end{equation}
These guarantee the unitarity of the matrix $A_f$ within the 
present experimental bound \cite{pdg} since 
$|B_f|, |C_f|\sim \frac{|m_f|}{|{\cal F}_f|} < O(10^{-7})$ is satisfied in each component. 
In such a case, it is easy to find 
\begin{equation}
A_f^{-1}\tilde m_f^2A_f\simeq m_fm_f^\dagger -\frac{1}{\mu_F^2
+{\cal F}_f{\cal F}_f^\dagger}
(m_f{\cal F}_f^\dagger)({\cal F}_fm_f^\dagger), \quad
\tilde M_F^2\simeq \mu_F^2+ {\cal F}_f{\cal F}_f^\dagger.
\label{ckm}
\end{equation}
The right-hand side of the first equation corresponds to an effective squared mass matrix 
of the ordinary fermions $f$. It is derived through the mixing with the extra heavy fermions $F$. 
Since its second term can have complex phases in off-diagonal 
components as long as $y_{f_i}\not=\tilde y_{f_i}$ is satisfied, 
the matrix $A_f$ could be complex. Moreover, if $\mu_F^2~{^<_\sim}~{\cal F}_f{\cal F}_f^\dagger$
is realized, the complex phase of $A_f$ in eq.~(\ref{ckm}) could have a substantial magnitude
because the second term in the right-hand side has a comparable magnitude with the first one.
Although it can be realized for various parameter settings, we consider 
a rather simple situation here,\footnote{Leptogenesis 
could depend on the strength of these couplings heavily in this model as studied later. }   
\begin{equation}
\langle\phi\rangle \ll w < u, \qquad  y_{f_j}\sim \tilde y_{f_j} < y_F.
\label{uw}
\end{equation}
Since the masses of vector-like fermions are expected to be of $O(10^8)$~GeV
or larger in this case, they decouple from the SM and do not contribute to 
phenomena at TeV regions.

The CKM matrix is determined as $V_{\rm CKM}={O_u}^TA_d$, where $O_u$ is an
orthogonal matrix used for the diagonalization of a mass matrix for up-type quarks.
Thus, the $CP$ phase of $V_{\rm CKM}$ is caused by the one of $A_d$.
The same argument is applied to the leptonic sector and
the PMNS matrix is derived as $V_{\rm PMNS}=A_e^\dagger U_\nu$, where 
$U_\nu$ is an orthogonal matrix used for the diagonalization 
of a neutrino mass matrix.
The Dirac $CP$ phase in the CKM matrix and the PMNS matrix can be 
induced from the same origin of $CP$ violation. 
A concrete example of $A_d$ is given for a simple case in Appendix A. 

\subsection{Effective model at a lower energy region}
An effective model at lower energy regions than $w$ and 
$u$ can be obtained by integrating
out the heavy fields. It is reduced to the SM with a lepton sector extended as 
the scotogenic neutrino mass model \cite{ma}, 
which is characterized by the terms invariant
under the remaining $Z_2$ symmetry
\begin{eqnarray}
-{\cal L}_{\rm scot}&=&\sum_{\alpha=1}^3
\left[\sum_{j=1}^3\left(\tilde h_{\alpha j}^\ast\bar\ell_\alpha\eta N_j 
+\frac{M_{N_j}}{2}\bar N_j^cN_j \right)
+ \sum_{\beta=1}^3\frac{y_{\alpha\beta}}{M_\ast}(\bar\ell_\alpha\phi)(\bar\ell_\beta\phi)+     
{\rm h.c.}\right] \nonumber \\
&+&\tilde m_\phi^2\phi^\dagger\phi+\tilde m_\eta^2\eta^\dagger\eta+
\tilde\lambda_1(\phi^\dagger\phi)^2+ \tilde\lambda_2(\eta^\dagger\eta)^2
+\tilde\lambda_3(\phi^\dagger\phi)(\eta^\dagger\eta)
+\lambda_4(\phi^\dagger\eta)(\eta^\dagger\phi) \nonumber \\
&+&\frac{\tilde\lambda_5}{2}\left[(\phi^\dagger\eta)^2+{\rm h.c.}\right],
\label{model0}
\end{eqnarray}
where neutrino Yukawa couplings $\tilde h_{\alpha j}$ are defined on the basis
for which the mass matrix of the charged leptons is diagonalized as discussed
in the previous part. Thus, they are complex now.
After the spontaneous breaking due to the VEVs of $\tilde\sigma$ and $\tilde S$,
coupling constants in eq.~(\ref{model0}) are related to the ones contained in eq.~(\ref{model}) as
\begin{eqnarray}
&&\tilde\lambda_1=\lambda_1-\frac{\kappa_{\phi\sigma}^2}{4\tilde\kappa_\sigma}
-\frac{\kappa_{\phi S}^2}{4\tilde\kappa_S}+
\frac{\kappa_{\sigma S}\kappa_{\phi\sigma}
\kappa_{\phi S}}{4\tilde\kappa_\sigma\tilde\kappa_S}, \qquad
\tilde\lambda_2=\lambda_2-\frac{\kappa_{\eta\sigma}^2}{4\tilde\kappa_\sigma}
-\frac{\kappa_{\eta S}^2}{4\tilde\kappa_S}+
\frac{\kappa_{\sigma S}\kappa_{\eta\sigma}
\kappa_{\eta S}}{4\tilde\kappa_\sigma\tilde\kappa_S}, \nonumber\\
&&\tilde\lambda_3=\lambda_3-\frac{\kappa_{\phi\sigma}\kappa_{\eta\sigma}}
{2\tilde\kappa_\sigma}
-\frac{\kappa_{\phi S}\kappa_{\eta S}}{2\tilde\kappa_S}+
\frac{\kappa_{\sigma S}\kappa_{\phi\sigma}\kappa_{\eta S}
+\kappa_{\sigma S}\kappa_{\eta\sigma}\kappa_{\phi S}}{4\tilde\kappa_\sigma\tilde\kappa_S},
\qquad \tilde\lambda_5=\lambda_5\frac{w}{M_\ast}. 
\end{eqnarray}
These connection conditions should be imposed
at a certain scale $\bar M$, which is taken to be $\bar M=\tilde M_F$ in the present study. 
Stability of the potential (\ref{model0}) requires the following conditions 
to be satisfied
through scales $\mu<\bar M$:
\begin{equation}
\tilde\lambda_1,~\tilde\lambda_2 >0, \qquad \tilde\lambda_3,~
\tilde\lambda_3+\lambda_4-|\tilde\lambda_5|>
-2\sqrt{\tilde\lambda_1\tilde\lambda_2}.
\label{stab2}
\end{equation} 
Potential stability (\ref{stab1}) and (\ref{stab2}) and perturbativity
of the model from the weak scale to the Planck scale can be examined by using
renormalization group equations (RGEs) for the coupling constants.
Relevant RGEs at $\mu>\bar M$ are given in Appendix B. 
The mass parameters in eq.~(\ref{model0}) are represented as 
\begin{eqnarray} 
&&M_{N_j}=y_{N_j}w,\nonumber\\
&&\tilde m_\phi^2=m_\phi^2+\left(\kappa_{\phi\sigma}+
\frac{\kappa_{\phi S}\kappa_{\sigma S}}{2\tilde\kappa_S}\right)w^2
+\left(\kappa_{\phi S}+
\frac{\kappa_{\phi\sigma}\kappa_{\sigma S}}{2\tilde\kappa_\sigma}\right)u^2, \nonumber\\
&&\tilde m_\eta^2=m_\eta^2+\left(\kappa_{\eta\sigma}+
\frac{\kappa_{\eta S}\kappa_{\sigma S}}{2\tilde\kappa_S}\right)w^2
+\left(\kappa_{\eta S}+
\frac{\kappa_{\eta\sigma}\kappa_{\sigma S}}{2\tilde\kappa_\sigma}\right)u^2.
\label{cpara}
\end{eqnarray}
If $\tilde m_\eta^2>0$ is satisfied and $\eta$ has no VEV, $Z_2$ is kept as an exact 
symmetry of the model. In this model, we assume both $|\tilde m_\phi|$ and $\tilde m_\eta$ 
have values of $O(1)$~TeV.  
Since it has to be realized under the contributions from the VEVs $w$ and $u$, 
parameter tunings are required.\footnote{For parameter values 
assumed in this analysis, the order of required tuning is estimated as $O(10^{-10})$.}

\begin{figure}[t]
\begin{center}
\includegraphics[width=7cm]{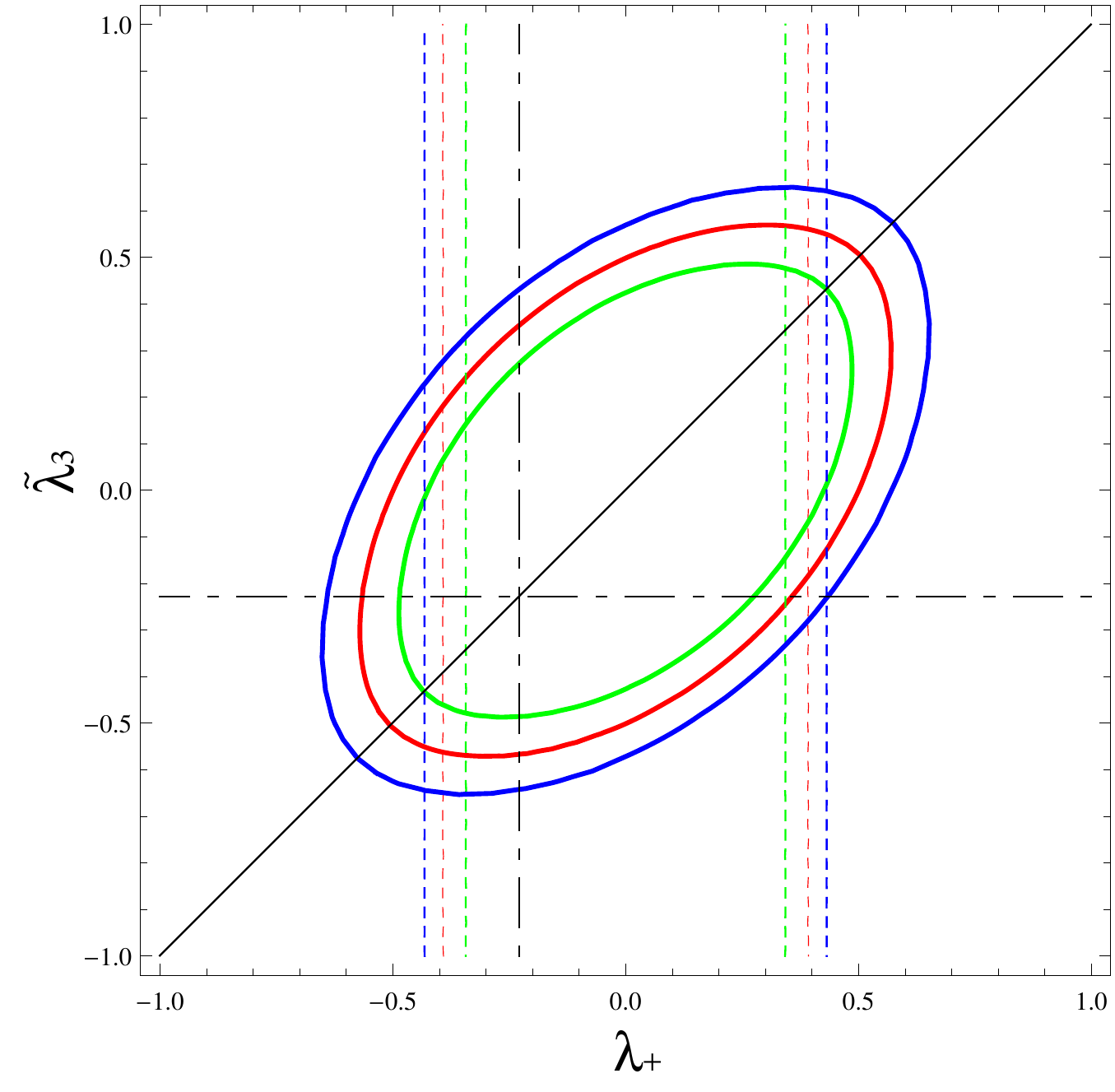}
\hspace*{5mm}
\includegraphics[width=7cm]{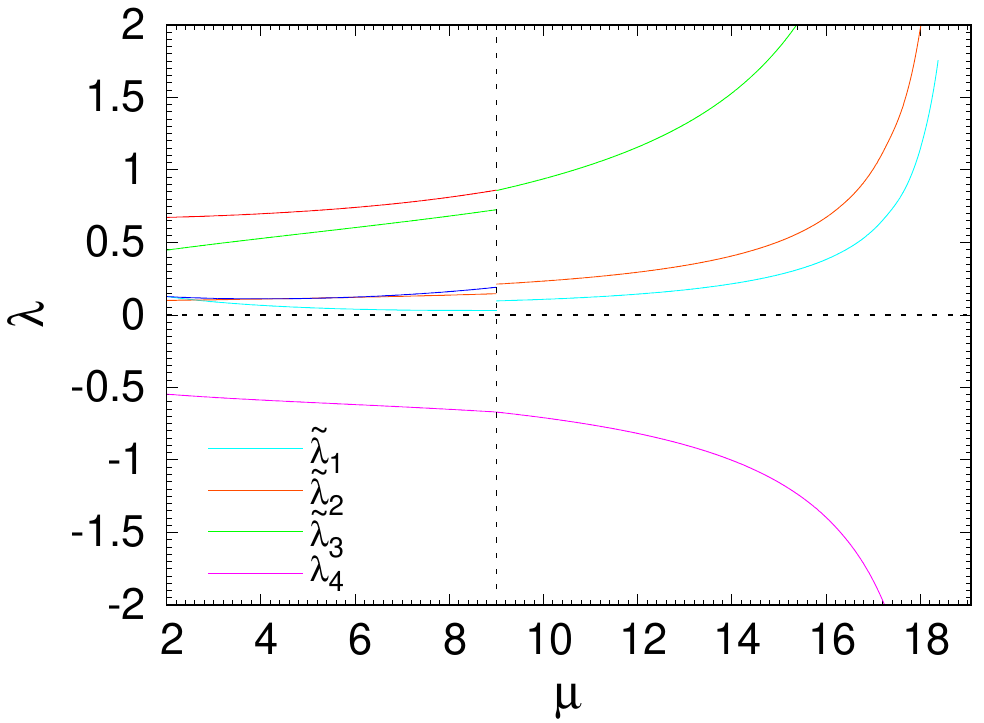}
\end{center}

\footnotesize{{\bf Fig.~1}~~Left:~contours of $\Omega h^2=0.12$ are plotted in the 
$(\lambda_+,\tilde\lambda_3)$ plane
by a solid line for $M_{\eta_R}=0.9$ (green), 1 (red) and 1.1 (blue) in a TeV unit.
Since $\eta_R$ should be lighter than the charged components, $\lambda_4<0$ should be satisfied
which corresponds to a region above the diagonal black solid line.  
Direct search bound from Xenon1T \cite{xenon} is shown by the same color 
dashed line for each $M_{\eta_R}$. 
Stability conditions (\ref{stab2}) restrict the allowed region to the area marked in the upper
 right quadrant (marked off by the dot-dashed lines), for which $\tilde\lambda_2=0.1$ is assumed. 
Right:~an example of the running of coupling constants $\tilde\lambda_i$ for
$M=10^\mu~{\rm GeV}$. Initial values of $\tilde\lambda_3$ and $\lambda_4$ are fixed as $\tilde\lambda_3=0.445$ and 
$\lambda_4=-0.545$, which are included in the allowed region of the left panel 
for $M_{\eta_R}=1$~TeV.  $\tilde\lambda_3+2\sqrt{\tilde\lambda_1\tilde\lambda_2}$ and
$\lambda_++2\sqrt{\tilde\lambda_1\tilde\lambda_2}$ 
are also plotted by red and blue lines, respectively.}
\end{figure}

Phenomenology on neutrinos and DM could be the same 
as the one which has been studied extensively in various studies \cite{scot,ks} 
unless the axion is a dominant component of DM. 
If the lightest neutral component of $\eta$ is DM which is identified as
its real component $\eta_R$, both DM
relic abundance and DM direct search constrain the parameters $\tilde\lambda_3$ and 
$\lambda_4$ \cite{mksvz,hs}. As a reference, in Fig.~1 we show their required values
in the $(\lambda_+,\tilde\lambda_3)$ plane for the cases 
$M_{\eta_R}=$0.9, 1 and 1.1 TeV where $\lambda_+=\tilde\lambda_3+\lambda_4+\tilde\lambda_5$
and $M_{\eta_R}^2=m_\phi^2+\lambda_+\langle\phi\rangle^2$.
They should be also consistent with the stability condition (\ref{stab2}).
The figure shows that these could be satisfied with rather restricted values of $\tilde\lambda_3$ 
and $\lambda_4$. A perturbativity requirement at $\mu>\bar M$ also constrains the model 
strongly since DM relic abundance requires $\tilde\lambda_3$ and $|\lambda_4|$ 
to take rather large values. We have to take account of it to consider the model 
at high energy regions. As an example, we show the result of RGE study in the
right panel of Fig.~1. Details of the assumed initial values are given in Appendix B.
If the initial values of $\tilde\lambda_3$ and $\lambda_4$ are chosen 
at a larger $\tilde\lambda_3$ region in the left panel,
the perturbatibity condition is violated at a much lower region than the Planck scale. 
We do not plot the behavior of $\kappa_i$ in this figure since their values are fixed to  be very 
small and they do not change their values substantially. 
If they satisfy the condition (\ref{stab1}), it can be confirmed to be kept up to the Planck scale.

Neutrino mass is forbidden at tree level due to this $Z_2$ symmetry except for
the ones generated through dimension five Weinberg operators in eq.~(\ref{model0}).
They could give a substantial contribution to the neutrino masses depending on 
the cut-off scale $M_\ast$ and coupling constants $y_{\alpha\beta}$. 
In the present study, however, we assume that their contribution is
negligible and the dominant contribution comes from one-loop diagrams with 
$\eta$ and $N_j$ in internal lines. Its formula is given as
\begin{equation}
{\cal M}^\nu_{\alpha\beta}\simeq\sum_{j=1}^3\tilde h_{\alpha j}\tilde 
h_{\beta j}\tilde\lambda_5\Lambda_j,  \qquad
\Lambda_j=\frac{\langle\phi\rangle^2}
{8\pi^2}\frac{1}{M_{N_j}}\ln\frac{M_{N_j}^2}{M_\eta^2},
\label{nmass}
\end{equation}
where $M_\eta^2=\tilde m_\eta^2+(\tilde\lambda_3+
\lambda_4)\langle\phi\rangle^2$
and $M_{N_j}\gg M_\eta$ is supposed.
As an example, one may assume a simple flavour 
structure for neutrino Yukawa couplings \cite{tribi}
\begin{equation}
\tilde h_{e i}=0, ~\tilde h_{\mu i}=\tilde h_{\tau i}\equiv h_i~~(i=1,~2); \quad  
\tilde h_{e 3}=\tilde h_{\mu 3}=-\tilde h_{\tau 3}\equiv h_3. 
\label{tribi}
\end{equation}
This realizes a tri-bimaximal mixing which gives a simple and rather good 0-th order 
approximation for the analysis of neutrino oscillation data and leptogenesis \cite{ks}. 
If we impose the mass eigenvalues obtained from eq.~(\ref{nmass}) for the case
$|h_1|\ll |h_2|, |h_3|$ to satisfy the squared mass differences required by the neutrino 
oscillation data, we find    
\begin{equation}
|h_2^2\tilde\lambda_5|\Lambda_2=\frac{1}{2}\sqrt{\Delta m_{32}^2}, \qquad 
|h_3^2\tilde\lambda_5|\Lambda_3=\frac{1}{3}\sqrt{\Delta m_{21}^2}.
\label{oscilcon}
\end{equation}
Since we have $\Lambda_{2,3}\sim7\times 10^5$~eV for $M_{2,3}\sim 10^{7}$~GeV
and $M_\eta\sim 10^3$~GeV, 
the neutrino oscillation data \cite{pdg} can be explained by taking as an example
\begin{equation}
y_{N_j}\sim 10^{-2}, \quad  |h_2|\sim 6\times 10^{-3},  \quad  |h_3|\sim 2\times 10^{-3}, 
\quad |\tilde\lambda_5|\sim 10^{-3}.
\label{npara}
\end{equation}
Even if we impose the neutrino oscillation data, $h_1$ can take a very small value 
compared with $h_{2,3}$ \cite{ks}. 
It can play a crucial role for low scale leptogenesis as seen later.

\section{Inflation due to singlet scalars}
\subsection{Inflation}
It is well known that a scalar field which couples nonminimally with the Ricci scalar 
can cause inflation of the universe
and the idea has been applied to the Higgs scalar in the SM \cite{higgsinf} and 
its singlet scalar extensions \cite{sinfl,ext-s}.
If the singlet scalars $S$ and $\sigma$, which are related to the $CP$ 
issues in the SM couple with the Ricci scalar, 
it can play the role of inflaton in this model. 
The action relevant to the inflation is given in the Jordan frame as
\begin{eqnarray}
S_J &=& \int d^4x\sqrt{-g} \left[-\frac{1}{2}M_{\rm pl}^2R -
\xi_\sigma\sigma^\dagger\sigma R-
\xi_{S1} S^\dagger S R -\frac{\xi_{S2}}{2}(S^2+S^{\dagger 2})R\right. \nonumber \\
&+&\left.\partial^\mu \sigma^\dagger \partial_\mu \sigma
+\partial^\mu S^\dagger \partial_\mu S - V(\sigma, S) \right],
\label{inflag}
\end{eqnarray}
where $M_{\rm pl}$ is the reduced Planck mass and the coupling of $S$ 
is controlled by the $Z_4$ symmetry.
$V(\sigma, S)$ stands for the corresponding part in the potential (\ref{model}).
Since inflation follows very complicated dynamics if multi scalars contribute to it,
we confine our study to the inflation in a potential valley.
Moreover,  we assume $\xi_{S1}=-\xi_{S2}$ is satisfied for simplicity. 
In that case, the coupling of $S$ with the Ricci scalar is reduced to $\frac{1}{2}\xi_S S_I^2R$, 
where $S=\frac{1}{\sqrt 2}(S_R+iS_I)$ and $\xi_S=\xi_{S1}-\xi_{S2}$.
If $S$ is supposed to evolve along a constant $\rho$, which is determined as 
a potential minimum $\frac{\partial V_b}{\partial\rho}=0$, the radial component $\tilde S$ 
couples with the Ricci scalar as $\frac{1}{2}\tilde\xi_S\tilde S^2R$, 
where $\tilde\xi_S$ is defined as $\tilde\xi_S=\xi_S\sin^2\rho$ and
the potential $V(\sigma, S)$ is expressed by eq.(\ref{potv}). 
Here we consider cases such that both $\xi_\sigma$ and $\tilde\xi_S$ are 
positive only.
Stability of this potential requires the condition given in eq.~(\ref{stab1}).
We neglect the VEVs $w$ and $u$ for a while since they are
much smaller than $O(M_{\rm pl})$, which is the field values of $\tilde\sigma$ and $\tilde S$ 
during the inflation. We also suppose that other scalars have much smaller values 
than them.

We consider the conformal transformation for a metric tensor in the Jordan frame
\begin{equation}
\tilde g_{\mu\nu}=\Omega^2g_{\mu\nu}, \qquad
\Omega^2=1+\frac{(\xi_\sigma\tilde\sigma^2+\tilde\xi_S\tilde S^2)}{M_{\rm pl}^2}.
\end{equation}
After this transformation to the Einstein frame where the Ricci scalar term takes a canonical form,
the action can be written as
\begin{eqnarray}
S_E &=& \int d^4x\sqrt{-\tilde g} \Big[-\frac{1}{2}M_{\rm pl}^2\tilde R
+\frac{1}{2}\partial^\mu\chi_\sigma \partial_\mu \chi_\sigma+\frac{1}{2}
\partial^\mu\chi_S \partial_\mu \chi_S \nonumber\\
&+&\frac{6\xi_\sigma\tilde\xi_S\frac{\tilde\sigma\tilde S}{M_{\rm pl}^2}}
{\left[(\Omega^2+\frac{6\xi_\sigma^2}{M_{\rm pl}^2}\tilde\sigma^2)
(\Omega^2+\frac{6\tilde\xi_S^2}{M_{\rm pl}^2}\tilde S^2)\right]^{1/2}}
\partial^\mu\chi_\sigma\partial_\mu\chi_S
- \frac{1}{\Omega^4}V(\tilde\sigma, \tilde S) \Big],
\label{inflag}
\end{eqnarray}
where $\chi_\sigma$ and $\chi_S$ are defined as \cite{sinfl}
\begin{equation}
\frac{\partial\chi_\sigma}{\partial\tilde\sigma}
=\frac{1}{\Omega^2}\sqrt{\Omega^2+6\xi_\sigma^2
\frac{\tilde\sigma^2}{M_{\rm pl}^2}}, \qquad
\frac{\partial\chi_S}{\partial\tilde S}
=\frac{1}{\Omega^2}\sqrt{\Omega^2+6\tilde\xi_S^2
\frac{\tilde S^2}{M_{\rm pl}^2}}.
\label{canoninf}
\end{equation} 
If we introduce variables $\tilde\chi,~\varphi$ to express $\tilde\sigma$ and $\tilde S$ as
$\tilde\sigma=\tilde\chi\cos\varphi,~ \tilde S=\tilde\chi\sin\varphi$, 
the potential in the Einstein frame 
at the large field regions
such as $\xi_\sigma\tilde\sigma^2+\tilde\xi_S\tilde S^2\gg M_{\rm pl}^2$
can be written as
 \begin{equation}
V(\tilde\chi,\varphi)=\frac{M_{\rm pl}^4}{4}\frac{\tilde\kappa_S\sin^4\varphi+
\tilde\kappa_\sigma\cos^4\varphi+\kappa_{\sigma S}
\sin^2\varphi\cos^2\varphi}{(\xi_\sigma\cos^2\varphi+\tilde\xi_S\sin^2\varphi)^2}.
\label{vk}
\end{equation} 
We find that there are three types of valley along the minimum in the $\varphi$
direction of this potential. They realize different types of inflaton. Two of them are 
\begin{equation}
{\rm (i)}~\varphi=0  ~~ {\rm for}~~ 2\tilde\kappa_\sigma\tilde\xi_S<
\kappa_{\sigma S}\xi_\sigma, \qquad
{\rm (ii)}~\varphi=\frac{\pi}{2} ~~ {\rm for} ~~ 2\tilde\kappa_S\xi_\sigma<
\kappa_{\sigma S}\tilde\xi_S.
\end{equation}
In each case, a kinetic term mixing between $\chi_\sigma$ and $\chi_S$ 
disappears and inflaton can be identified with 
$\chi_\sigma$ for (i) and $\chi_S$ for (ii), 
respectively.\footnote{In different context, the inflaton dominated by $\tilde\sigma$
and the $\tilde S$ inflaton have been discussed in 
\cite{smash} and \cite{patisalam}, respectively.} 

Another valley which is studied in this paper is realized at 
\begin{equation}
\sin^2\varphi=\frac{2\tilde\kappa_\sigma\tilde\xi_S-\kappa_{\sigma S}\xi_\sigma}
{(2\tilde\kappa_S\xi_\sigma-\kappa_{\sigma S}\tilde\xi_S)
+(2\tilde\kappa_\sigma\tilde\xi_S-\kappa_{\sigma S}\xi_\sigma)},
\label{varphi}
\end{equation}
under the conditions
\begin{equation}
 2\tilde\kappa_\sigma\tilde\xi_S>\kappa_{\sigma S}\xi_\sigma, \quad 
2\tilde\kappa_S\xi_\sigma>\kappa_{\sigma S}\tilde\xi_S,
\end{equation} 
where we note that these are automatically satisfied for $\kappa_{\sigma S}<0$ 
since eq.~(5) is imposed, and $\tilde\xi_S$ and $\xi_\sigma$ are assumed to be positive. 
In this case the inflaton $\tilde\chi$ is a mixture of $\tilde\sigma$ and $\tilde S$ 
with a constant value of $\tilde\sigma/\tilde S$.
Although the kinetic term mixing cannot be neglected for a general $\sin\varphi$, 
it can be safely neglected if we restrict it to the one in which the inflaton is dominated 
by $\tilde S~(\sin^2\varphi\simeq 1)$ or $\tilde\sigma~(\sin^2\varphi\simeq 0)$.
We focus our study on the former case where 
$\tilde\chi\gg\tilde\sigma$ is always satisfied during inflation.
If we additionally impose $\tilde\xi_S \gg \xi_\sigma$ on eq.~(\ref{varphi})
and assume that the relevant couplings satisfy\footnote{If we assume 
$\tilde\kappa_S>|\kappa_{\sigma S}|$, $\sin\varphi$ is differently expressed as   
$\sin^2\varphi=1-\frac{\tilde\kappa_S\xi_\sigma}{\tilde\kappa_\sigma\tilde\xi_S}$.
However, we do not consider such a case in this paper.}
\begin{equation}
\kappa_{\sigma S}<0, \qquad \tilde\kappa_S< |\kappa_{\sigma S}|< \tilde\kappa_\sigma,
\label{ks}
\end{equation}
$\sin\varphi$ is expressed as 
$\sin^2\varphi=1+\frac{\kappa_{\sigma S}}{2\tilde\kappa_\sigma}$.
In this case, by using $\hat\kappa_S=\tilde\kappa_S-\frac{\kappa_{\sigma S}^2}
{4\tilde\kappa_\sigma}$ potential can be expressed as $V(\tilde S)=\frac{1}{4\Omega^4}
\hat\kappa_S\tilde S^4$ at the bottom of the valley and 
$\cos 2\rho=-\frac{\beta}{4\alpha}\cot^2\varphi$ realizes the minimum of $V_b$ if
$\tan^2\varphi>\left|\frac{\beta}{4\alpha}\right|$ is satisfied. 
Nature of the inflaton $\tilde\chi$ is fixed by the parameters $\tilde\kappa_S$, 
$\tilde\kappa_\sigma$ and $\kappa_{\sigma S}$.

Squared mass of the orthogonal component to $\tilde\chi$ during inflation can be 
estimated as $m_{\tilde\chi_\perp}^2=\frac{|\kappa_{\sigma S}|M_{\rm pl}^2}{2\tilde\xi_S^2}$. 
Since the Hubble parameter $H_I$ satisfies 
$H_I^2=\frac{\hat\kappa_SM_{\rm pl}^2}{12\tilde\xi_S^2}$ at the same period,
$H_I < m_{\tilde\chi_\perp}$ is satisfied under the condition (\ref{ks}).\footnote{ 
Mass of another orthogonal component $S_I$
is estimated as $m_{S_I}^2\sim \frac{8\alpha M_{\rm pl}^2}{\tilde\xi_S^2}$ for a case 
$\cos 2\rho\sim 0$, which is interested in the present study
and then $96\alpha >\hat\kappa_S$ is required for $m_{S_I} >H_I$.
Although $\alpha \ll \hat\kappa_S$ is assumed from the vacuum determination, 
it can be satisfied for suitable values of $\alpha$.}
 Thus, the inflation $\chi$ starts rolling along the valley within a few Hubble time 
independently of an initial value of the inflaton. It justifies our analyzing the model as 
a single field inflation model.
On the other hand, since $\tilde\sigma(=\sqrt\frac{|\kappa_{\sigma S}|}
{2\tilde\kappa_\sigma}\tilde\chi)>\frac{H_I}{\sqrt{2\pi}}$ is satisfied
generally,\footnote{Although it may be escapable for 
$\tilde\xi_S\ll\sqrt{|\kappa_{\sigma S}|}$, it is not the case in the present model.}
the global $U(1)$ is spontaneously broken during inflation and isocurvature 
fluctuation could be problematic \cite{smash}. 
However, even in that case it is escapable since the axion 
needs not to be a dominant component of DM in the present model. 
This problem is discussed later.

The canonically normalized inflaton $\chi$ can be expressed as \cite{smash}
\begin{equation}
\Omega^2\frac{d\chi}{d\tilde S}=\sqrt{\gamma\Omega^2+6\tilde\xi_S^2
\frac{\tilde S^2}{M_{\rm pl}^2}},  
\label{tchi}
\end{equation}
where $\gamma$ can be approximated along the valley as 
\begin{equation}
\gamma=1-\frac{\kappa_{\sigma S}}{2\tilde\kappa_\sigma}.
\end{equation}
If we use $\gamma\simeq 1$, the potential of $\chi$ obtained through
$V(\chi)=\frac{1}{\Omega^4}V(\tilde\sigma,\tilde S)$
can be derived by using the solution of eq.~(\ref{tchi}) which is given as
\begin{equation}
\frac{\chi}{M_{\rm pl}}=-\sqrt 6~{\rm arcsinh}\left(\frac{\sqrt\frac{6}{\gamma}
\frac{\tilde\xi_S \tilde S}{M_{\rm pl}}}
{\sqrt{1+\frac{\tilde\xi_S}{M_{\rm pl}^2}\tilde S^2}}\right)+
\sqrt\frac{\gamma+6\tilde\xi_S}{\tilde\xi_S}{\rm arcsinh}
\left(\frac{\sqrt{\tilde\xi_S(1+\frac{6}{\gamma}\tilde\xi_S)}\tilde S}{M_{\rm pl}}\right).
\label{etchi}
\end{equation}
We derive the potential of $\chi$ through numerical calculation for a 
typical value of $\tilde\xi_S$ by using eq.~(\ref{etchi}). 
Such examples of $V(\chi)$ are shown in Fig.~2.
It can be approximated as 
\begin{eqnarray}
V(\chi)=\left\{ \begin{array}{ll}
\frac{\hat\kappa_S}{4\tilde\xi_S^2}M_{\rm pl}^4 &\quad \chi>M_{\rm pl}, \\
\frac{\hat\kappa_S}{6\tilde\xi_S^2}M_{\rm pl}^2\chi^2  
&\quad \frac{M_{\rm pl}}{\tilde\xi_S}
<\chi< M_{\rm pl}, \\
 \frac{\hat\kappa_S}{4}\chi^4 &\quad \chi < \frac{M_{\rm pl}}{\tilde\xi_S}.
\end{array}\right.
\label{oscilpot}
\end{eqnarray}
The inflation ends at $\chi_{\rm end}\simeq M_{\rm pl}$.
After the end of inflation, there is a substantial region where the potential
behaves as a quadratic form before it is reduced to a quartic form 
at low energy regions for the case $\tilde\xi_S\gg 1$ as in the Higgs inflation. 
However, such a region can be neglected for the case $\tilde\xi_S<10$,
which is the case considered in this study. 
Since the inflaton oscillating in the quartic potential behaves 
as radiation as shown later, the radiation domination starts 
soon after the end of inflation in that case.

\begin{figure}[t]
\begin{center}
\includegraphics[width=7.5cm]{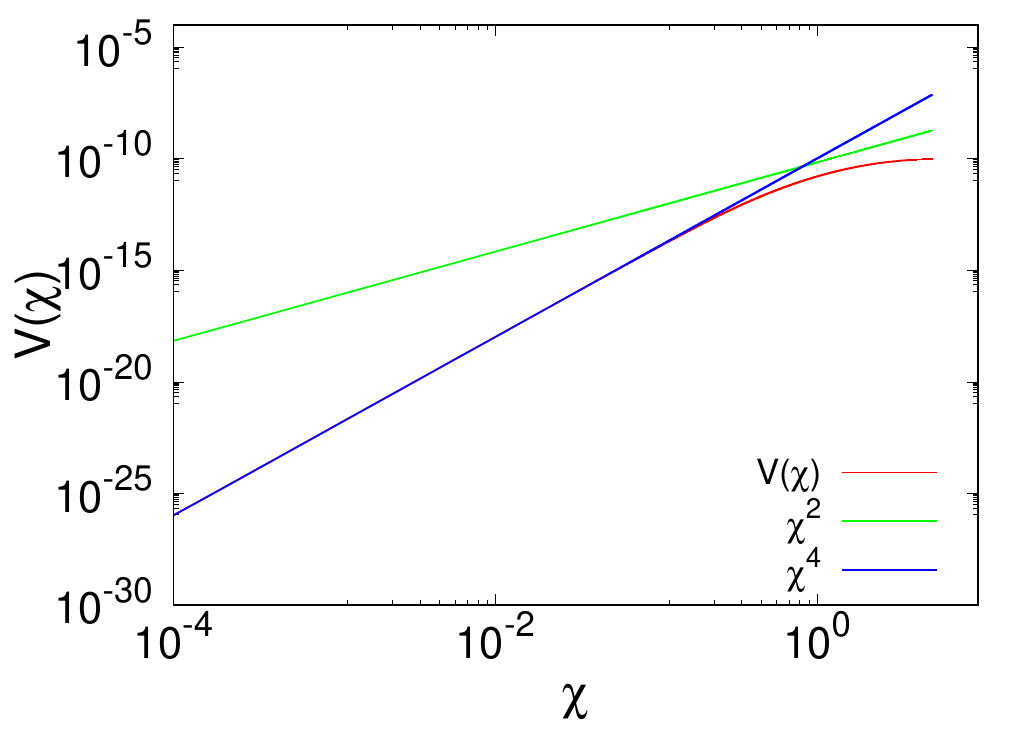} 
\hspace*{3mm}
\includegraphics[width=7.5cm]{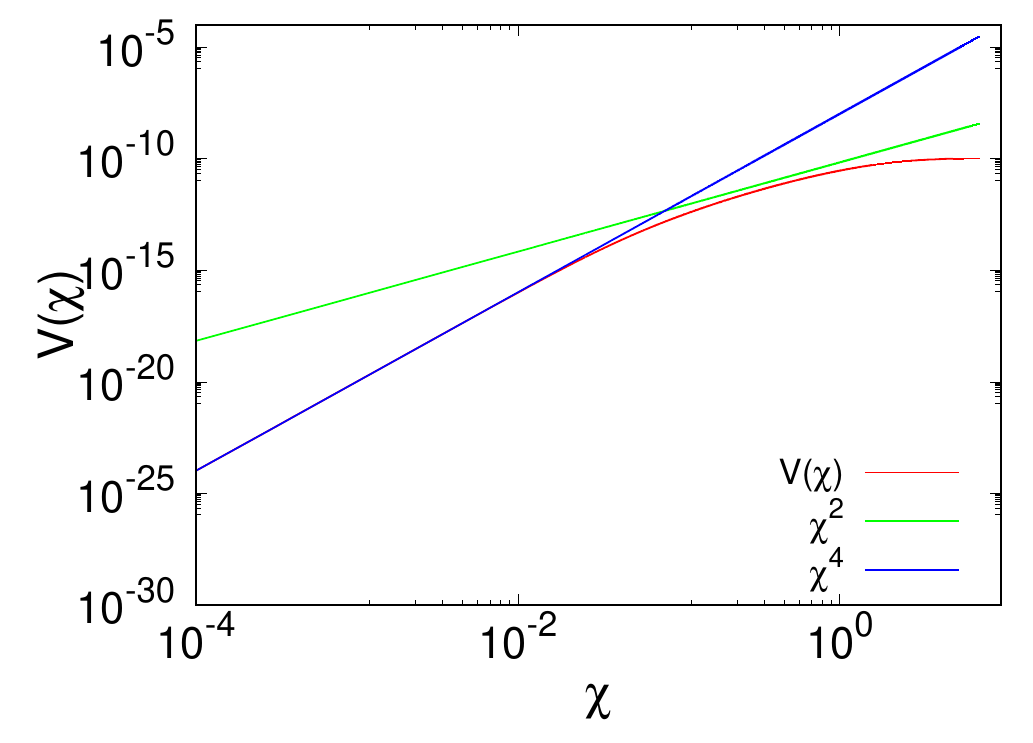}
\end{center}

\footnotesize{{\bf Fig.~2}~~Potential of the inflaton $\chi$ for 
$\tilde\xi_S=1$ (left panel) and $\tilde\xi_S=10$ (right panel).
In both panels, $\tilde\xi_S/\xi_\sigma=20$ and 
$\tilde\kappa_S/|\kappa_{\sigma S}|=|\kappa_{\sigma S}|/\tilde\kappa_\sigma=0.1$ 
are assumed and $\tilde\kappa_S$ is fixed by using 
eq.~(\ref{kap}) for ${\cal N}_k=55$. As references, we also plot
approximated potential $\frac{\hat\kappa_S}{6\tilde\xi_S^2}M_{\rm pl}^2\chi^2$ 
and $\frac{\hat\kappa_S}{4}\chi^4$ in eq.~(\ref{oscilpot}) as $\chi^2$ and $\chi^4$.
In these plots, a Planck unit $(M_{\rm pl}=1)$ is used.}
\end{figure}

The slow-roll parameters in this model can be estimated by using 
eq.~(\ref{tchi}) as \cite{inf}
\begin{equation}
\epsilon\equiv\frac{M_{\rm pl}^2}{2}\left(\frac{V^\prime}{V}\right)^2
=\frac{8M_{\rm pl}^4}{\gamma\tilde\xi_S\left(1+\frac{6}{\gamma}\tilde\xi_S\right)\tilde S^4}, \qquad
\eta\equiv M_{\rm pl}^2\frac{V^{\prime\prime}}{V}=
-\frac{8M_{\rm pl}^2}{\gamma\left(1+\frac{6}{\gamma}
\tilde\xi_S\right)\tilde S^2}.
\end{equation}
The $e$-foldings number ${\cal N}_k$ from the time when the scale 
$k$ exits the horizon to the end of inflation is estimated by using eq.~(\ref{tchi}) as 
\begin{eqnarray}
{\cal N}_k=\frac{1}{M_{\rm pl}^2}\int^{\chi_k}_{\chi_{\rm end}}\frac{V}{V^\prime}d\chi
=\frac{1}{8M_{\rm pl}^2}(\gamma+6\tilde\xi_S)(\tilde S_k^2-\tilde S_{\rm end}^2)
-\frac{3}{4}\ln\frac{M_{\rm pl}^2+\tilde\xi_S\tilde S_k^2}
{M_{\rm pl}^2+\tilde\xi_S\tilde S_{\rm end}^2}.
\label{efold}
\end{eqnarray}
Taking account of these,
the slow-roll parameters in this inflation scenario is found to be approximated as 
$\epsilon\simeq \frac{3}{4{\cal N}_k^2}$ and $\eta\simeq -\frac{1}{{\cal N}_k}$.
The field value of inflaton during the inflation is found to be  expressed as 
$\chi_k=\frac{\sqrt 6}{2}M_{\rm pl}\ln(32\tilde\xi_S{\cal N}_k)$
by using eqs.~(\ref{etchi}) and (\ref{efold}), and its potential $V_k(\equiv V(\chi_k))$ takes 
a constant value as shown in eq.~(\ref{oscilpot}).
On the other hand, if we use $\epsilon=1$ at the end of inflation,
 the inflaton potential is estimated as 
$V_{\rm end}(\equiv V(\chi_{\rm end}))\simeq 0.072
\frac{\hat\kappa_S}{\tilde\xi_S^2}M_{\rm pl}^4$ which is found to be a 
good approximation from Fig.~2. 

The spectrum of density perturbation predicted by the inflation 
is known to be expressed as \cite{inf}
\begin{equation}
{\cal P}(k)=A_s\left(\frac{k}{k_\ast}\right)^{n_s-1},  \qquad
A_s=\frac{V}{24\pi^2M_{\rm pl}^4\epsilon}\Big|_{k_\ast}. 
\label{power}
\end{equation}
If we use the Planck data $A_s=(2.101^{+0.031}_{-0.034})\times 10^{-9}$ 
at $k_\ast=0.05~{\rm Mpc}^{-1}$ \cite{planck18}, we find the
Hubble parameter during the inflation to be $H_I=1.4\times 10^{13}
\left(\frac{60}{{\cal N}_{k_\ast}}\right)$~GeV and the relation 
\begin{equation}
\hat\kappa_S\simeq 4.13\times 10^{-10}\tilde\xi_S^2 \left(\frac{60}{{\cal N}_{k_\ast}}\right)^2,
\label{kap}
\end{equation}
which should be satisfied at the horizon exit time of the scale $k_\ast$.
We confine our study to the case $\tilde\xi_S<10$.

\begin{figure}[t]
\begin{center}
\includegraphics[width=8cm]{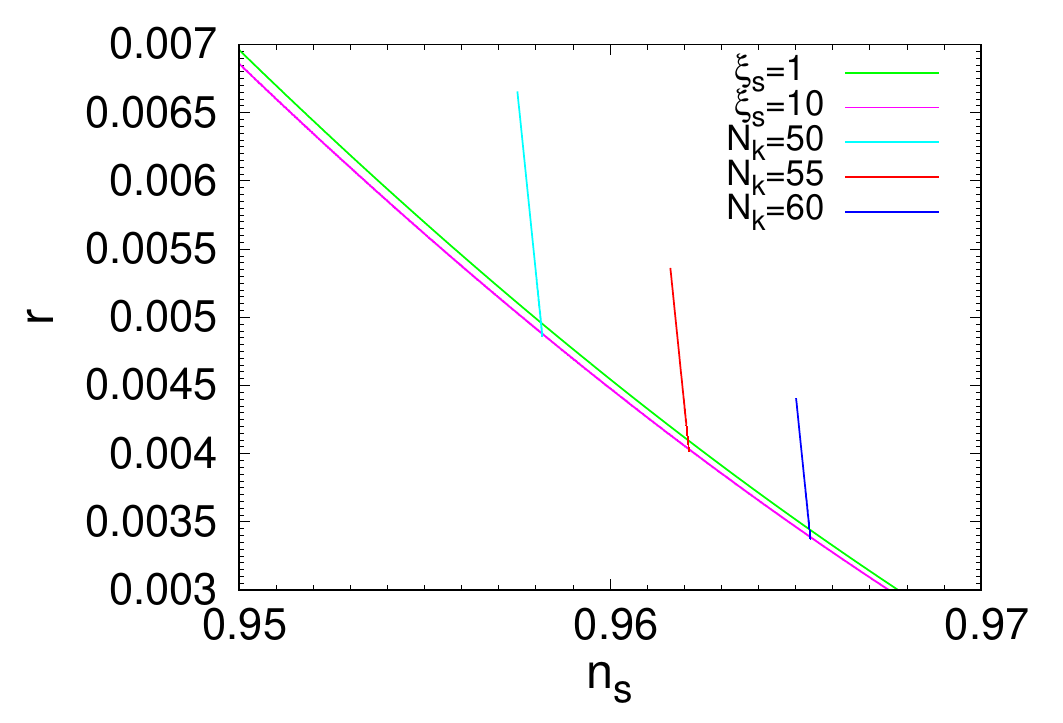} 
\end{center}

\footnotesize{{\bf Fig.~3}~~Predicted values of the scalar spectral index $n_s$ 
and the tensor-to-scalar ratio $r$ in this model. They are read off as the values 
at intersection points of two lines with a fixed value of $\tilde\xi_S$ or ${\cal N}_k$. 
The coupling constant $\hat\kappa_S$ is varied in a range 
from $10^{-7}$ to $10^{-10}$.}
\end{figure}

In Fig.~3, we plot predicted values  
for the scalar spectral index $n_s$ and the tensor-to-scalar ratio $r$ 
in the present model.
Since the quartic coupling $\hat\kappa_S$ is a free parameter of the model
under the constraint (\ref{kap}),
we vary $\hat\kappa_S$ in the range $10^{-10}\le \hat\kappa_S\le 10^{-7}$
for fixed values of $\tilde\xi_S$ or ${\cal N}_k$.
The CMB constraint (\ref{kap}) is satisfied at intersection points of 
the lines with a fixed value of $\tilde\xi_S$ or ${\cal N}_k$.
The figure shows that the constraints of the observed CMB data 
\cite{planck18} are satisfied for the supposed parameters. 

After the end of inflation, the inflaton $\chi$ starts oscillation in the potential $V(\chi)$.
At this stage, the description by $\chi$ is no longer justified, especially, at 
the small field regions. $\varphi$ is not constant in general there. $\tilde S$ and $\tilde\sigma$
should be treated independently. In the following study, however, 
we confine our study to a special inflaton trajectory and estimate the reheating
phenomena by using $\chi$ to give a rough evaluation of reheating temperature 
under the assumption that inflaton follows a constant $\varphi$ trajectory.\footnote{
During the first several oscillations, both $\varphi$ and $\rho$ can 
be numerically confirmed to take constant values. 
This suggests that single field treatment is rather good 
during the first few oscillations at least.} 
In this case, inflaton oscillation is described by the equation
\begin{equation}
\frac{d^2\chi}{dt^2}+3H\frac{d\chi}{dt}+V^\prime(\chi)=0.
\label{infeq}
\end{equation}
Since the amplitude of $\chi$ evolves approximately as  
$\Phi(t)=\frac{\tilde\xi_S}{\sqrt{\pi\hat\kappa_S}t}$ in the quadratic 
potential after the end of inflation, the inflaton $\chi$ oscillates 
$\frac{1}{2\pi\sqrt{3\pi}}(\tilde\xi_S-1)$ times before the potential (\ref{oscilpot}) 
changes from a quadratic form to a quartic one. 
This means that preheating under the quadratic potential could play 
no substantial role for the case $\tilde\xi_S<10$. 
In such a case, we need to consider the preheating in the quartic potential only. 
The model with the quartic potential $V(\chi)=\frac{\hat\kappa_S}{4}\chi^4$
becomes conformally invariant \cite{confpre}. 

If we introduce dimensionless conformal time $\tau$, which is defined 
by using a scale factor $a$ as $a d\tau=\sqrt{\hat\kappa_S}\chi_{\rm end}dt$
and also a rescaled field $f=\frac{a\chi}{ \chi_{\rm end}}$,
eq.~(\ref{infeq}) can be rewritten as
\begin{equation}
\frac{d^2f}{d\tau^2}+f^3=0.
\label{infeq1}
\end{equation}  
The solution of this equation which describes the inflaton oscillation is known 
to be given by a Jacobi elliptic function 
$f(\tau)={\rm cn} \left(\tau-\tau_i,\frac{1}{\sqrt 2}\right)$.\footnote{If we take 
$\tau_i\simeq 2.44$, $f(\tau)$ can be approximated by $\cos\left(\frac{2\pi}{\tau_0}\tau\right)$,
where $\tau_0$ is expressed by using the complete elliptic integral of the first kind $K$
as $\tau_0=4K(\frac{1}{\sqrt 2})$ \cite{confpre}.}
From the Friedman equation for this inflaton oscillation, we find 
\begin{equation}
a(\tau)=\frac{\chi_{\rm end}}{2\sqrt3 M_{\rm pl}}\tau, \qquad  
\tau=2(3\hat\kappa_S M_{\rm pl}^2)^{1/4}\sqrt t.
\label{rd}
\end{equation} 
Since $H=1/2t$ is satisfied, this oscillation era is radiation dominated.
If we take into account such a feature of the model that radiation domination 
starts just after the end of inflation, the $e$-foldings number ${\cal N}_k$ 
can be expressed by noting a relation $k=a_kH_k$ as
\begin{equation}
{\cal N}_k=56.7-\ln\left(\frac{k}{a_0H_0}\right)+
\ln\left(\frac{V_{\rm end}^{1/4}}{10^{14}~{\rm GeV}}\right)
+2\ln\left(\frac{V_k^{1/4}}{V_{\rm end}^{1/4}}\right),
\label{efolds}
\end{equation} 
where $H_k^2=\frac{V_k}{3M_{\rm pl}^2}$ and the suffix 0 stands for the present 
value of each quantity. Reheating temperature dependence of ${\cal N}_k$ is weak or lost differently 
from the usual case \cite{inf} where substantial matter domination is assumed 
to follow the inflation era.\footnote{If reheating occurs through a perturbative process
at $\chi~{^<_\sim}~u$ where matter domination is realized, its effect on ${\cal N}_k$ 
could also be negligible as long as $\Gamma>H$ is satisfied at that stage where
$\Gamma$ is inflaton decay width.}  
In the next part, we discuss the reheating temperature expected to be 
realized in the present model. 

\subsection{Preheating and reheating}
Before proceeding to the study of particle production under the background
oscillation of inflaton, we need to know the mass of the relevant particles 
which is induced through the interaction with inflaton.
Such interactions are given as
\begin{eqnarray}
&&\Big[-\frac{y_D}{\sqrt 2}\frac{\kappa_{\sigma S}}{2\tilde\kappa_\sigma}
\chi\bar D_LD_R 
-\frac{y_E}{\sqrt 2}\frac{\kappa_{\sigma S}}{2\tilde\kappa_\sigma}
\chi\bar E_LE_R+
\sum_{j=1}^3\Big\{
\frac{1}{\sqrt 2}\left(y_{d_j} e^{i\rho} 
+ \tilde y_{d_j} e^{-i\rho}\right)\chi\bar D_L d_{R_j} \nonumber\\ 
&&+\frac{1}{\sqrt 2}\left(y_{e_j} e^{i\rho} + 
\tilde y_{e_j} e^{-i\rho}\right)\chi\bar E_L e_{R_j}
-\frac{y_{N_j}}{2\sqrt 2}\frac{\kappa_{\sigma S}}{2\tilde\kappa_\sigma}
\chi \bar N_j^cN_j \Big\}
+ {\rm h.c.}\Big] \nonumber \\
&&+\frac{1}{2}(\kappa_{\phi S}\phi^\dagger\phi+\kappa_{\eta S}\eta^\dagger\eta)
\chi^2
-\frac{\kappa_{\sigma S}}{2\tilde\kappa_\sigma}
(\kappa_{\phi\sigma}\phi^\dagger\phi
+\kappa_{\phi\sigma}\eta^\dagger\eta)\chi^2.
\label{oscm}
\end{eqnarray} 
The particles interacting with the inflaton $\chi$ have mass varying 
with the oscillation of $\chi$ and their mass can be read off from 
eq.~(\ref{oscm}) as
\begin{eqnarray}
&&M_{N_j}\simeq \frac{y_{N_j}}{\sqrt 2}\frac{|\kappa_{\sigma S}|}
{2\tilde\kappa_\sigma}\chi, \quad
\tilde M_{F}\simeq \frac{\chi}{\sqrt 2}\left[\sum_{j=1}^3(y_{f_j}^2+\tilde y_{f_j}^2) +
y_F^2\frac{\kappa_{\sigma S}^2}
{4\tilde\kappa_\sigma^2}\right]^{\frac{1}{2}}, \nonumber \\
&&m_\phi^2\simeq\frac{1}{2}\left(\kappa_{\phi S}+
\frac{|\kappa_{\sigma S}|}{\tilde\kappa_\sigma}\kappa_{\phi\sigma}\right)\chi^2, 
\quad 
m_\eta^2\simeq\frac{1}{2}\left(\kappa_{\eta S}+
\frac{|\kappa_{\sigma S}|}{\tilde\kappa_\sigma}
\kappa_{\eta\sigma}\right)\chi^2,
\label{imass1}
\end{eqnarray}
where $F=D$ or $E$ should be understood for $f=d$ or $e$, respectively.
Since the effect of nonminimal coupling is negligible during this oscillation period, 
it is convenient to use the components of $\sigma$ and $S$, which are 
parallel and orthogonal to the inflaton $\chi$ to describe their interactions.
If we indicate each of them as $\sigma_\parallel$, $\sigma_\perp$, 
$S_\parallel$, and $S_\perp$, then their interactions are expressed as 
\begin{eqnarray}
&&\frac{\tilde\kappa_S}{4}(S_\parallel^2+S_\perp^2-u^2)^2
+\frac{\kappa_{\sigma S}}{4}(S_\parallel^2+S_\perp^2-u^2)
(\sigma_\parallel^2+\sigma_\perp^2-w^2) 
+\frac{\tilde\kappa_\sigma}{4}\left(\sigma_\parallel^2
+\sigma_\perp^2-w^2\right)^2.\nonumber \\
\end{eqnarray}  
By combining these interactions with the composition of $\chi$,
their masses are found to be given by\footnote{It should be noted 
that the mass of $\sigma$
could have another non-negligible contribution which is
induced by explicit breaking of the global $U(1)$ symmetry 
brought about by the quantum gravitational effects.
We do not take account of it in the present study.}
\begin{eqnarray}
&& m^2_{S_\parallel}\simeq \left(3\hat\kappa_S+
\frac{\kappa_{\sigma S}^2}{2\tilde\kappa_\sigma}\right)\chi^2, \qquad
m_{\sigma_\parallel}^2\simeq \left(|\kappa_{\sigma S}|+
\frac{\kappa_{\sigma S}^2}{4\tilde\kappa_\sigma}\right)\chi^2, \nonumber \\
&&m^2_{S_\perp}\simeq\hat\kappa_S \chi^2, \qquad
m_{\sigma_\perp}^2\simeq
\frac{\kappa_{\sigma S}^2}{4\tilde\kappa_\sigma}\chi^2.
\label{imass2}
\end{eqnarray}

The coupling constants relevant to these masses are restricted 
through the assumed inflaton composition and the realization of the 
$CP$ phases in the CKM and PMNS matrices.  
The discussion in the previous sections shows that such requirements 
are satisfied for
\begin{equation} 
\hat\kappa_S < |\kappa_{\sigma S}|<\tilde\kappa_\sigma, \qquad 
\hat\kappa_S< y_{N_j}, y_{f_j}, \tilde y_{f_j}.
\label{rehpara}
\end{equation}
We assume additionally 
\begin{equation}
\hat\kappa_S < g_\phi\equiv \kappa_{\phi S}+
\frac{|\kappa_{\sigma S}|}{\tilde\kappa_\sigma}\kappa_{\phi\sigma},\qquad
\hat\kappa_S < g_\eta\equiv \kappa_{\eta S}+
\frac{|\kappa_{\sigma S}|}{\tilde\kappa_\sigma}\kappa_{\eta\sigma}.
\label{pecoupl}
\end{equation}
Since the oscillation frequency of the inflaton is 
$\sim \sqrt{\hat\kappa_S}\chi$, decays or annihilations of the inflaton are 
kinematically forbidden except for the one to $\sigma_\perp$ 
as found from eqs.~(\ref{imass1}) and (\ref{imass2}). 
In $\sigma_\perp$ case, the inflaton reaction rate to it is much smaller than the 
Hubble parameter at this period because of the smallness of its coupling with the inflaton, 
energy drain from the inflaton to $\sigma_\perp$ is ineffective to be neglected. 
As a result, the energy transfer from the inflaton oscillation to excited particles 
is expected to occur at the time when the inflaton crosses the zero where 
the resonant particle production is possible.

Preheating under the background inflaton oscillation can generate the excitations of 
$\chi$ itself and other scalars $\psi$ which couple with $\chi$ 
at its zero crossing \cite{pre}.
In a quartic potential case \cite{confpre}, the model becomes 
conformally invariant and 
the time evolution equations of $\chi_k(\simeq S_k)$ 
and $\psi_k$, which are the comoving modes with a momentum $k$,  
 can be transformed to the simple ones by rescaling them 
to the dimensionless quantities 
in the  same way as eq.~(\ref{infeq1}). They are given as 
\begin{eqnarray}
&& \frac{d^2}{d\tau^2}X_k+\omega_k^2X_k=0, \qquad 
\omega_k^2=\bar k^2+3f(\tau)^2, \nonumber \\
&& \frac{d}{d\tau^2}F_k+\tilde\omega_k^2F_k=0, \qquad 
\tilde\omega_k^2=\bar k^2+\frac{g_\psi}{\hat\kappa_S}f(\tau)^2, 
\label{teq}
\end{eqnarray}
where the rescaled variables are defined as
\begin{eqnarray}
X_k=\frac{a\chi_k}{\chi_{\rm end} }, \qquad 
F_k=\frac{a\psi_k}{\chi_{\rm end} }, \qquad 
\bar k=\frac{ak}{\chi_{\rm end}\sqrt{\hat\kappa_S}}.
\end{eqnarray}
Function $f(\tau)$ is the solution of eq.~(\ref{infeq1}) and $g_\psi$ stands 
for a coupling constant of the relevant particle 
$\psi(=\sigma, S_{\perp},\phi,\eta)$ with 
the inflaton $\chi$, which can read off from eqs.~(\ref{imass1}) and (\ref{imass2}). 
Amplitudes $X_k$ and $F_k$ are known to show the exponential behavior
$\propto e^{\mu_k\tau}$ with a characteristic exponent $\mu_k$, which is
determined by a parameter $g_\psi/\hat\kappa_S$.
Using the solutions of eq.~(\ref{teq}), the number density
of the produced particle $\psi$ can be calculated as
\begin{equation}
n_k^\psi=\frac{\tilde\omega_k}{2\hat\kappa_S}
\left(\frac{|F_k^\prime|^2}{\tilde\omega_k^2} +|F_k|^2\right)-\frac{1}{2}.
\end{equation}

Particle production based on eq.~(\ref{teq}) at the inflaton zero crossing 
has been studied in \cite{confpre} and it is shown to be characterized by 
the parameter $g_\psi/\hat\kappa_S$.
We classify the relevant couplings into five groups  
\begin{equation}
{\rm (A)}~\frac{g_{\sigma_\parallel}}{\hat\kappa_S}\gg 1, \quad 
{\rm (B)}~\frac{g_{S_\parallel}}{\hat\kappa_S}=3, \quad 
{\rm (C)}~\frac{g_{S_\perp}}{\hat\kappa_S}=1, \quad
{\rm (D)}~\frac{g_{\sigma_\perp}}{\hat\kappa_S} \ll 1,\quad
{\rm (E)}~\frac{g_\phi}{\hat\kappa_S},~\frac{g_\eta}{\hat\kappa_S}>1,
\end{equation}
where we note that couplings in (A)-(D) are fixed by the present 
inflaton composition but the ones in (E) are not constrained.
Now we consider the resonant particle production in each group. 
A maximum value of characteristic exponent in (D) is very small so that it plays no effective
role also in preheating.
In (B) and (C), both the fluctuations of $S_\parallel$ 
and $S_\perp$ are produced fast, but it stops as soon as $\langle |S_\parallel|^2\rangle$ 
and $\langle |S_\perp|^2\rangle$ reach a certain value such as $0.5\chi_{\rm end}^2/a^2$.
Although a maximum value $\mu_{\max}$ of the characteristic exponent of (B) is much 
smaller than the one of (C) and also the resonance band of (B) is much narrower than (C),
the interaction $S_\parallel^2S_\perp^2$ accelerates the production of fluctuations 
of $S_\parallel$ through rescattering and they reach the similar value \cite{smash, rescat}.
Since the backreaction of these fluctuations to the inflaton oscillation restructures 
the resonance band,  the resonant particle production stops before causing 
much more conversion of the inflaton oscillation energy to particle excitations.
Moreover, since the decay of excitations produced through these processes are also closed
kinematically, these could not play an efficient role in reheating.
In (A), since $\sigma_\parallel$ also couples to the inflaton directly,  
the resonant production of its excitation stops at a certain stage due to the same 
reason as (B) and (C). Even if the excited particles are allowed to 
decay to fermions $F$ 
and $N_j$ kinematically, the decay width is much smaller than 
the Hubble parameters to be neglected.
As a result, if the process due to (E) is not effective, 
preheating cannot play any role for reheating and reheating proceeds
through perturbative processes after the amplitude of inflaton is smaller 
than the VEV $u$. 

Here, we have to note that there is a possibility in (E) where the energy 
transfer from the inflaton oscillation to radiation proceeds through preheating 
since the produced excitations can decay to relativistic particles differently from (B) and (C).
In this case, $\phi$ and $\eta$ are produced as excitations at the zero crossing 
of the inflaton where an adiabaticity condition $\tilde\omega_k^\prime<\tilde\omega_k^2$ 
could be violated for certain values of $\bar k$. 
By using the analytic solution of eq.~(\ref{teq}) derived in \cite{confpre}, 
the momentum distribution $n_{k}^\psi$ of the produced particle $\psi$ through 
one zero crossing of the inflaton can be estimated as 
\begin{equation}
n_{\bar k}^\psi= e^{2\mu_k\frac{\tau_0}{2}}=e^{-\left(\bar k/\bar k_c\right)^2},
\qquad \bar k_c^2=\sqrt\frac{g_\psi}{2\pi^2\hat\kappa_S},
\end{equation}
where $\tau_0$ is an inflaton oscillation period and $\tau_0=7.416$. 
The resonance is efficient for $\bar k<\bar k_c$.
Thus, the particle number density produced during one zero crossing of the 
inflaton is   
\begin{equation}
\bar n^\psi =\int\frac{d^3\bar k }{(2\pi)^3}n_{\bar k}^\psi
=\int\frac{d^3\bar k}{(2\pi)^3}e^{-(\bar k/\bar k_c)^2}
=\frac{\bar k_c^3}{8\pi^{3/2}}.
\end{equation}

The energy transfer from the inflaton oscillation to relativistic particles
is caused through the decay of the produced particles $\psi(=\phi, \eta)$ 
and thermalization proceeds.
They can decay to light fermions through $\phi\rightarrow \bar qt$ 
with a top Yukawa coupling $h_t$ 
and $\eta \rightarrow \bar\ell N$ with neutrino Yukawa couplings $h_j$, respectively. 
Here, we should note that $\eta$ can be heavier than $N_j$ at this 
stage even if $\eta$ 
is the lightest one with $Z_2$ odd parity at the weak scale.
It is caused by the inflaton composition in the present model 
as found from eq.~(\ref{imass1}).  
Their decay widths in the comoving frame are given 
by using the conformally rescaled unit as
\begin{eqnarray}
&&\bar\Gamma_\psi=\frac{c_\psi y_\psi^2}{8\pi}\bar m_\psi,  \qquad 
\bar m_\psi=\frac{a m_\psi}{ \chi_{\rm end} \sqrt{\hat\kappa_S}}
=\sqrt\frac{g_\psi}{\hat\kappa_S}f(\tau), 
\end{eqnarray}
where $\psi=\phi,~\eta$, and $c_\psi$ are internal degrees 
of freedom $c_\phi=3$ and $c_\eta=1$.
The Yukawa coupling $y_\psi$ represents $y_\phi=h_t$ and $y_\eta=h_j$. 
Since $\bar\Gamma_{\psi}^{-1}<\tau_0/2$ is satisfied for $g_\psi> 4\times 
10^{-7}\left(\frac{\hat\kappa_S}{10^{-8}}\right)$, the produced $\psi$
decays to the light fermions completely before the next inflaton 
zero crossing \cite{inst} and then it is not accumulated in such cases. 
We fix $\tau=0$ at the first inflaton zero crossing so that $f(\tau)$ can be
expressed approximately as $f(\tau)=f_0\sin(cf_0\tau)$.
Transferred energy density through the $\psi$ decay during a half period of 
oscillation can be estimated as\footnote{$\bar\rho$ is 
defined as the energy density in the comoving frame 
by using the conformally rescaled variables.}
\begin{equation}
\delta\bar\rho_r=\int^{\tau_0/2}_0d\tau\bar\Gamma_\psi\bar m_\psi\bar n_\psi
e^{-\int_0^\tau\bar\Gamma_\psi\tau^\prime}
=\frac{1}{8\pi^{3/2}(2\pi^2)^{3/4}}\left(\frac{g_\psi}{\hat\kappa_S}\right)^{5/4}
Y(f_0,\gamma_\psi),
\end{equation}
where $\gamma_\psi$ and $Y(f_0,\gamma_\psi)$ are defined by using 
$c=2\pi/\tau_0$ as
\begin{equation}
\gamma_\psi=\frac{c_\psi y_\psi^2}{8\pi c}\sqrt\frac{g_\psi}{\hat\kappa_S}, \qquad
Y(f_0,\gamma_\psi)=c\gamma_\psi\int^{\tau_0/2}_0d\tau f_0^2\sin^2(cf_0\tau)
e^{-2\gamma_\psi\sin^2(\frac{cf_0\tau}{2})}.
\end{equation}

The energy density transferred to the light particles is accumulated at each 
inflaton zero crossing linearly and its averaged value for $\tau$ is 
estimated as
\begin{equation}
\bar\rho_r(\tau)=\frac{2\tau}{\tau_0}\delta\bar\rho_r
=6.5\times 10^{-4}\left(\frac{g_\psi}{\hat\kappa_S}\right)^{5/4}
Y(f_0,\gamma_\psi)\tau,
\label{rde}
\end{equation}
where the substantial change of $f_0$ is assumed to be negligible
during $\tau$.
Since the total energy density of the inflaton oscillation energy $\bar\rho_\chi$
and the transferred energy $\bar\rho_r$ to light particles  
is conserved, reheating temperature realized through
this process can be estimated from $\bar\rho_{\chi_{\rm end}}=\bar\rho_r$. 
It can be written by transferring it to the physical unit as
\begin{equation}
\frac{1}{4\hat\kappa_S}\left(\frac{\sqrt{\hat\kappa_S}
\chi_{\rm end}}{a}\right)^4=\frac{\pi^2}{30}g_\ast T_R^4,
\end{equation}
where we use $\bar\rho_{\chi_{\rm end}}=\frac{1}{4\hat\kappa_S}$ and $g_\ast=130$.
By applying eqs.~(\ref{rd}) and (\ref{rde}) to this formula, 
we find\footnote{The same result can be obtained by using the relation 
$H=\frac{1}{2t}$ in the radiation dominated era together 
with eqs.~(\ref{rd}) and (\ref{rde}).}
\begin{equation}
T_R=5.9\times 10^{15}g_\psi^{5/4}Y(f_0,\gamma_\psi)~{\rm GeV}.
\label{nonptr}
\end{equation}
Since $h_t\gg h_j$ is satisfied, reheating temperature is expected to 
be determined by 
the produced $\phi$ as long as $\phi$ is dominantly produced.

If preheating cannot produce relativistic particles effectively, 
the dominant energy is still kept in the inflaton oscillation. 
When the oscillation amplitude of $\chi$ decreases to be $O(u)$, 
the inflaton starts decaying to the light particles through the perturbative processes.   
Since the mass pattern is expected under the present assumption
for the coupling constants in (\ref{rehpara}) to be
\begin{equation}
2\tilde m_\eta<m_\chi< \tilde M_D, \tilde M_E,
\label{mspec}
\end{equation}
the inflaton decay is expected to occur mainly through 
$\chi\rightarrow \eta^\dagger\eta$ and 
$\chi\rightarrow \phi^\dagger\phi$ at tree level.
The decay width of $\psi(=\phi,~\eta)$ is estimated as
\begin{eqnarray}
\Gamma_\psi&\simeq& 
\frac{g_\psi^2}{16\pi \hat\kappa_S}m_\chi, 
\label{width}
\end{eqnarray}
where $g_\psi$ is defined in eq.~(\ref{pecoupl}).
After the inflaton decays to $\eta^\dagger\eta$ and $\phi^\dagger\phi$,
the SM contents are expected to be thermalized through gauge interactions 
with $\eta$ and $\phi$ immediately. 
Since $\Gamma_\psi > H$ is satisfied for $g_\psi>10^{-7.1}
\left(\frac{\hat\kappa_S}{10^{-8}}\right)^{1/2}
\left(\frac{u}{10^{11}~{\rm GeV}}\right)^{1/2}$ at $\chi\simeq u$,
reheating temperature in such a case can be estimated through 
$\frac{1}{4}\hat\kappa_Su^4=\frac{\pi^2}{30}g_\ast T_R^4$ as\footnote{
Although a larger value of $u$ can make the reheating 
temperature much higher, its upper bound exits. 
Since a larger $g_\psi$ is required in that case,
 it could violate the perturbativity of the model and cause the upper bound for it. 
For example, if we consider the case with $y_F=10^{-1.2}$, 
the perturbativity is violated for $g_\psi>10^{-4.4}$. As a result, 
the reheating temperature due to the perturbative process 
is bounded as $T_R<6.3 \times 10^{13}$ GeV.}
\begin{equation}
T_R\simeq
2.8\times10^8\left(\frac{\hat\kappa_S}{10^{-8}}\right)^{1/4}
\left(\frac{u}{10^{11}~{\rm GeV}}\right)~{\rm GeV},
\end{equation}
which is independent of $g_\psi$. However, if $\Gamma_\psi > H$ is not satisfied 
because of a small $g_\psi$, the reheating temperature is expected to be determined 
through $\Gamma_\psi=H$ and then becomes smaller proportionally to $g_\psi$.

\begin{figure}[t]
\begin{center}
\includegraphics[width=7.5cm]{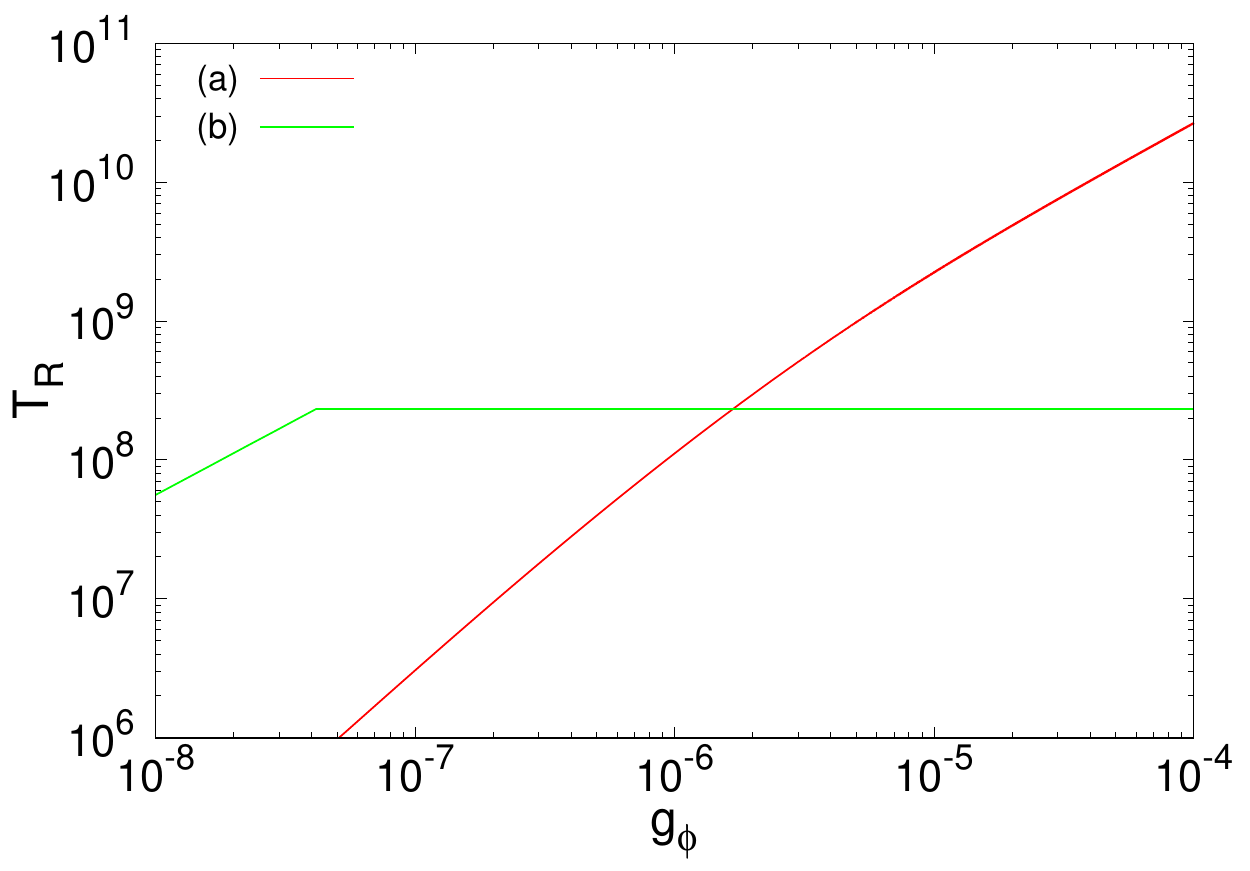} 
\hspace*{5mm}
\includegraphics[width=7.5cm]{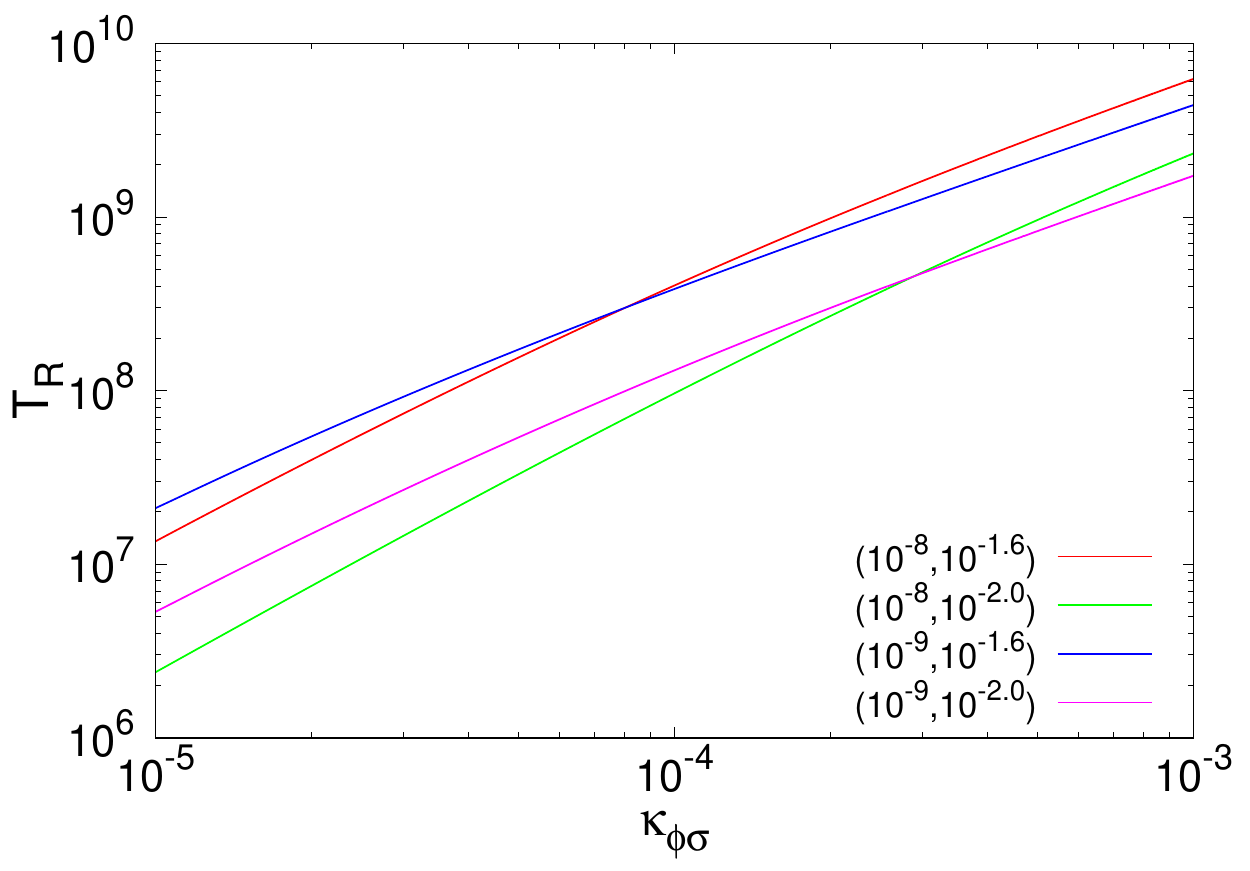} 
\end{center}

\footnotesize{{\bf Fig.~4}~~Left : reheating temperature $T_R$ (GeV) predicted
for both the preheating (a) and the perturbative process (b). 
They are plotted as a function of $g_\phi$.
In the case (a), the decay $\phi\rightarrow \bar qt$ is assumed. 
Right : contribution to the reheating temperature due to 
a component $\sigma$ of the inflaton in a case $\kappa_{\phi S}=0$. 
The reheating temperature $T_R$ (GeV) is plotted as a function of $\kappa_{\phi\sigma}$.
Each line represents $T_R$ for several values of 
$(\tilde\kappa_S, |\kappa_{\sigma S}|/\tilde\kappa_\sigma)$, where
$\tilde\kappa_\sigma$ is fixed at $10^{-4.5}$ and $10^{-5.3}$ for 
$\tilde\kappa_S=10^{-8}$ and $10^{-9}$, respectively.}
\end{figure}

In the left panel of Fig.~4, for a case $\psi=\phi$, 
the expected reheating temperature through both processes is plotted 
as a function $g_\phi$ in a case 
$\tilde\kappa_S=10^{-8}$, $\tilde\kappa_\sigma=10^{-4.5}$, 
$|\kappa_{\sigma S}|/\tilde\kappa_\sigma=10^{-1.2}$, and $u=10^{11}$ GeV.
It shows that the reheating temperature is determined by the perturbative 
process at $g_\phi<10^{-6}$. 
We also found from the figure that the reheating at $g_\phi>10^{-6}$ 
proceeds through the preheating. Even if the dominant component $S$ of
the inflaton has no coupling with $\phi$ so that 
$g_\phi\simeq|\kappa_{\sigma S}|\kappa_{\phi\sigma}/\tilde\kappa_\sigma$, 
the preheating is caused by the component $\sigma$.
It is shown in the right panel where the reheating temperature is plotted 
for $\kappa_{\phi\sigma}$ by varying 
$|\kappa_{\sigma S}|/\tilde\kappa_\sigma$.\footnote{The condition 
$\bar\Gamma_\psi^{-1}<\tau_0/2$ can be 
confirmed for the parameters used here. } 
These figures show that $T_R>2.3\times 10^8$ GeV can be realized 
if $g_\phi>4\times 10^{-8}$ is satisfied.
However, since the perturbativity of the model is found to be violated 
at $g_\phi>10^{-4.4}$ as mentioned in the previous footnote,
$\kappa_{\phi\sigma}<10^{-4.4}$ and $\kappa_{\phi\sigma}<10^{-2.6}$
should be satisfied, and then the reheating temperature cannot be 
higher than $\sim 10^{10}$~GeV as found from the figure.
Since the decay of $\phi$ is so effective, it decays soon 
after their production and much before the inflaton amplitude becomes large 
during the oscillation. This makes the energy transfer in the preheating  
inefficient.
 
In the usual leptogenesis in the seesaw scenario, the right-handed 
neutrinos are supposed to be thermalized only through the neutrino 
Yukawa couplings $h_j$.
In the present model, neutrino mass eigenvalues obtained from eq.~(\ref{nmass}) 
require $h_{2,3}=O(10^{-3})$ to explain the neutrino oscillation data 
as discussed at a part of eq.~(\ref{npara}).
On the other hand, reheating temperature is found to satisfy 
$T_R~{^>_\sim}~10^8$ GeV from the above discussion.
Since the decay width $\Gamma_{N_{2,3}}$ of $N_{2,3}$ and the reheating 
temperature $T_R$ satisfy $\Gamma_{N_{2,3}}>H(T_R)$
and $T_R>M_{N_{2,3}}$, $N_{2,3}$ are also expected to be 
in the thermal equilibrium through the inverse decay simultaneously 
at the reheating period.  
In the case of $N_1$, however, it depends on the magnitude of its Yukawa 
coupling $h_1$ which can be much smaller than others. 
We should note that $N_1$ could be effectively generated in the thermal bath,
even if $h_1$ is extremely small, through the scattering of extra fermions 
which are expected in the thermal equilibrium through gauge interactions 
in the case $\tilde M_D, \tilde M_E< T_R$.  
It is a noticeable feature of the present model which opens a window for 
low scale leptogenesis.

\section{Phenomenological signature of the model}
\subsection{Leptogenesis}
The most interesting feature of this inflation scenario is that 
thermal leptogenesis could generate sufficient baryon number asymmetry even 
for $M_{N_1}<10^9$ GeV without relying on resonance effect.
In the ordinary seesaw framework,
neutrino mass is generated as 
$(m_\nu)_{\alpha\beta}=\frac{h_{\alpha j}h_{\beta j}\langle\phi\rangle^2}{M_{N_j}}$ 
through Yukawa interaction $h_{\alpha j}\bar\ell_\alpha\phi N_j$.  
Baryon number asymmetry in the universe \cite{baryon} is 
expected to be generated by the same interaction through 
thermal leptogenesis \cite{leptg}. 
If we assume the sufficient lepton asymmetry is generated 
through the out-of-equilibrium decay of the lightest right-handed neutrino,
which has been in the thermal equilibrium, then
the reheating temperature $T_R$ is required to be larger than its mass 
$T_R>M_{N_1}$. 
Moreover, since it has to be produced sufficiently in the thermal bath, 
its Yukawa coupling $h_{\alpha 1}$ should not be so small. On the other hand, 
the neutrino mass formula gives a severer upper bound on $h_{\alpha 1}$ 
for a smaller $M_{N_1}$ under the constraints of neutrino oscillation data.
These impose a lower bound for $M_{N_1}$ such as $10^9$ GeV \cite{di}.  
This condition for $M_{N_1}$ is not changed even if $T_R\gg 10^9$ GeV is satisfied.
The problem is caused by such a feature of the model that both the production and 
the out-of-equilibrium decay of the right-handed neutrino have to be caused only 
by the same neutrino Yukawa coupling. 
It does not change in the original scotogenic model either \cite{ks}. 
In that model, the right-handed neutrino mass can be much smaller than $10^9$ GeV
keeping the neutrino Yukawa couplings to be rather larger values by fixing 
$|\lambda_5|$ at a smaller value in a consistent way with
the neutrino oscillation data. However, 
the washout of the generated lepton number due to the inverse decay of the right-handed
neutrinos becomes so effective in that case. 
As a result, successful leptogenesis cannot be realized
for a lighter right-handed neutrino than $10^8$ GeV.\footnote{Low scale 
leptogenesis in the scotogenic model has been studied intensively 
in \cite{lowlept1}. However, the lightest
right-handed neutrino is assumed to be in the thermal equilibrium initially there.}     
It is a notable aspect in the present model that this situation 
can be changed by the particles 
which are introduced to explain the $CP$ issues in the SM.

We note that the interaction between the right-handed 
neutrino $N_1$ and extra vector-like fermions $F$ mediated by $\tilde\sigma$
could change the situation.\footnote{The similar mechanism has been discussed 
in models with a different type of inflaton \cite{hs,lowlept2}.} 
The lightest right-handed neutrino $N_1$ can be effectively produced 
in the thermal bath through the extra fermions scattering 
$\bar D_LD_R, \bar E_LE_R\rightarrow N_1N_1$ 
mediated by $\tilde\sigma$ if $D_{L,R}$ and/or $E_{L,R}$ are in the thermal 
equilibrium at a certain temperature $T$.  In that case,
both conditions $T>\tilde M_F, M_{N_1}$ and $\Gamma_{FF}\simeq H(T)$
are required to be satisfied,
 where $\Gamma_{FF}$ is the reaction rate of this scattering. 
Mass of these fermions is determined by the VEVs $u$ and $w$ which should 
be larger than the lower bound of PQ symmetry breaking scale.
Since the rough estimation of  $\Gamma_{FF}\simeq H(T)$ for 
relativistic $F$ and $N_1$ gives 
\begin{equation}
T\simeq 5.8\times 10^8\left(\frac{y_F}{10^{-1.2}}\right)^2
\left(\frac{y_{N_1}}{10^{-2}}\right)^2 {\rm GeV},
\label{scat}
\end{equation}
we find that $T>\tilde M_F, M_{N_1}$ could be satisfied for suitable values of
$y_F$ and $y_{N_1}$.
It is crucial that this does not depend on the magnitude of 
the $N_1$ Yukawa coupling $h_1$. 
If an extremely small value is assumed for $h_1$,
successful leptogenesis is allowed in a consistent way with neutrino oscillation data
even for $M_{N_1}<10^9$ GeV. 

After $N_1$ is produced in the thermal bath through the scattering of
the extra fermions mediated by $\tilde\sigma$,
it is expected to decay to $\ell_\alpha\eta^\dagger$ by a strongly suppressed 
Yukawa coupling. 
Since its substantial decay occurs after the washout processes 
are frozen out, the generated lepton number asymmetry can be efficiently 
converted to the baryon number asymmetry through sphaleron processes.
This scenario can be checked by solving Boltzmann equations for $Y_{N_1}$ and 
$Y_L(\equiv Y_\ell-Y_{\bar\ell})$, where $Y_\psi$ is 
defined as $Y_\psi=\frac{n_\psi}{s}$
by using the $\psi$ number density $n_\psi$ and the entropy density $s$.
Boltzmann equations analyzed here are given as  
\begin{eqnarray}
&&\frac{dY_{N_1}}{dz}=-\frac{z}{sH(M_{N_1})}\left(\frac{Y_{N_1}}{Y^{\rm eq}_{N_1}}
-1\right)\left[ \gamma_D^{N_1}+ 
\left(\frac{Y_{N_1}}{Y^{\rm eq}_{N_1}}+1\right)\sum_{F=D,E}
\gamma_{F}\right], \nonumber \\
&&\frac{dY_{L}}{dz}=-\frac{z}{sH(M_{N_1})}\left[\varepsilon
\left(\frac{Y_{N_1}}{Y^{\rm eq}_{N_1}}-1\right)
\gamma_D^{N_1} -\frac{2Y_L}{Y_\ell^{\rm eq}}
\sum_{j=1,2,3}\left(\frac{\gamma_D^{N_j}}{4} +\gamma_{N_j}\right)\right],
\label{beq}
\end{eqnarray}
where $z=\frac{M_{N_1}}{T}$ and an equilibrium value of $Y_\psi$ is 
represented by $Y_\psi^{\rm eq}$.
$H(T)$ is the Hubble parameter at temperature $T$ and 
the $CP$ asymmetry $\varepsilon$ for the decay of $N_1$ is expressed as
\begin{eqnarray}
\varepsilon&=&\frac{1}{8\pi}\sum_{j=2,3}
\frac{{\rm Im}[ \sum_{\alpha}(\tilde h_{\alpha 1}
\tilde h_{\alpha j}^\ast)]^2}
{\sum_{\alpha} \tilde h_{\alpha 1}\tilde h_{\alpha 1}^\ast}
F\left(\frac{M_{N_j}^2}{M_{N_1}^2}\right)
\nonumber\\
&=&\frac{1}{16 \pi}\left[4|h_2|^2F\left(\frac{y_{N_2}^2}{y_{N_1}^2}\right)
\sin 2(\theta_1-\theta_2)+|h_3|^2F\left(\frac{y_{N_3}^2}{y_{N_1}^2}\right)
\sin 2(\theta_1-\theta_3)\right],
\label{epsilon}
\end{eqnarray}
where $h_j=|h_j|e^{i\theta_j}$ and $F(x)=\sqrt x\left[1-(1+x)\ln\frac{1+x}{x}\right]$.
A reaction density for the decay 
$N_j\rightarrow \ell_\alpha\eta^\dagger$ and for the lepton number violating 
scattering mediated by $N_j$ is expressed by $\gamma_D^{N_j}$ 
and $\gamma_{N_j}$, respectively \cite{ks}. 
$\gamma_{F}$ represents a reaction density for the scattering 
$\bar D_LD_R,\bar E_LE_R\rightarrow N_1N_1$. 
We assume that $(D_L,D_R)$ and $(E_L,E_R)$ are in the thermal equilibrium 
and $Y_{N_1}=Y_L=0$ at $z=z_R(\equiv\frac{M_{N_1}}{T_R})$. 

Now we fix the model parameters for numerical study of eq.~(\ref{beq}) by
taking account of the discussion in the previous part. 
We consider two cases for the VEVs of the singlet scalars such that 
\begin{equation}
{\rm (I)}~ w=10^9~{\rm GeV}, ~u=10^{11}~{\rm GeV},
\qquad  {\rm (II)}~ w=10^{11}~{\rm GeV},~ u=10^{13}~{\rm GeV},
\end{equation} 
where the axion could be a dominant DM in case (II).
The parameters $\tilde\kappa_S, \tilde\kappa_\sigma$, and $\kappa_{\sigma S}$, 
which characterize the inflaton $\chi$, are fixed to $\tilde\kappa_S=10^{-8}, 
\tilde\kappa_\sigma=10^{-4.5}$, and $|\kappa_{\sigma S}|=10^{-6.1}$. 
These are used in the right panel of  Fig.~4. 
The condition ${\cal F}_f{\cal F}_f^\dagger> \mu_F^2$ for which the
$CP$ phases in the CKM and PMNS matrices can be generated is 
reformulated as $\delta\equiv\tilde M_F/\mu_F>\sqrt 2$.  
If we confine our study on a case $y_f=(0,0,y)$ and $\tilde y_f=(0,\tilde y,0)$
for simplicity,\footnote{It is considered as an example in Appendix A.}
we have a relation $y^2+\tilde y^2=(\delta^2-1)\frac{w^2}{u^2}y_F^2$ 
among Yukawa couplings of the extra fermions.
We fix them as $\delta=\sqrt 3$, $\tilde y/y=0.5$ and $y_D=y_E=10^{-1.2}$ 
at the scale $\bar M$. 
Parameters relevant to the neutrino mass generation are fixed as
\begin{eqnarray}
&& y_{N_2}=2\times 10^{-2},  \quad  y_{N_3}=4\times 10^{-2}
\quad {\rm for ~(I)~ and~ (II)}, \nonumber\\ 
&& y_{N_1}=7\times 10^{-3}, \quad |h_1|=6\times 10^{-7} 
\quad |\tilde\lambda_5|=10^{-3}, \quad M_\eta=1~{\rm TeV}
\quad {\rm for~ (I)},
\nonumber \\
&&y_{N_1}= 10^{-3}, \quad |h_1|=6\times 10^{-5}, 
\quad |\tilde\lambda_5|=5\times 10^{-3}, \quad M_\eta=0.9~{\rm TeV}
\quad {\rm for~ (II)}.
\label{para}
\end{eqnarray}
These parameters fix the mass of relevant particles as
\begin{eqnarray}
{\rm (I)}&& M_{N_1}=7\times 10^6~{\rm GeV}, \quad M_{N_2}= 2\times 10^7~{\rm GeV}, \quad  
M_{N_3}=4\times 10^7~{\rm GeV}, \nonumber \\
&&\tilde M_D= \tilde M_E=1.1\times 10^8~{\rm GeV}, \quad m_{\tilde\sigma}=8\times 10^6~{\rm GeV},
\nonumber \\
{\rm (II)}&&M_{N_1}=10^8~{\rm GeV},  \quad M_{N_2}= 2\times 10^9~{\rm GeV}, \quad  
M_{N_3}=4\times 10^9~{\rm GeV}, \nonumber \\  
&&\tilde M_D= \tilde M_E=1.1\times 10^{10}~{\rm GeV}. \quad m_{\tilde\sigma}=8\times 10^8~{\rm GeV}. 
\end{eqnarray}
Although the mediator has a small component $\tilde S$, it can be safely 
treated as $\tilde\sigma$.
For these parameters, the $CP$ asymmetry $\varepsilon$ in the $N_1$ decay takes 
a value of $O(10^{-6})$ in both cases if the maximum $CP$ phase is assumed.
DM is determined by the couplings $\tilde\lambda_3$ and $\lambda_4$.
Since they are fixed so as to realize the correct DM abundance by 
the neutral component of $\eta$ in the case (I), it cannot saturate the required 
DM abundance for the same $\tilde\lambda_3$ and $\lambda_4$ in the case (II) 
as found from Fig.~1. The axion could be a dominant component of DM 
in the case (II) since $w$ is taken to be a sufficient value for it. 

We give a remark on these couplings here.
It is crucial to examine whether the above parameters used in this analysis are 
consistent with the potential stability conditions (\ref{stab1}), (\ref{stab2}) and 
also the perturbativity of the model under constraints coming 
from the requirements for the DM relic abundance and the reheating temperature.
If DM relic abundance is realized by the neutral component of $\eta$, 
both $\tilde\lambda_3$ and $\lambda_4$ should take values shown in Fig.~1. 
On the other hand, the reheating temperature required for sufficient 
leptogenesis can be realized for $\kappa_{\phi\sigma}~{^>_\sim}~10^{-4}$ 
or $\kappa_{\phi S}~{^>_\sim}~10^{-7}$ as found from the analysis of 
the reheating temperature. 
Since they can give rather large contributions to the $\beta$-functions 
of the scalar quartic couplings $\kappa_{\eta\sigma}$ and $\kappa_{\eta S}$
for example, the perturbativity up to the inflation scale could be violated. 
An upper bound on $g_\phi$ has to be imposed to escape it and
it results in an upper bound on the reheating temperature discussed already.  
The parameter sets used here have been confirmed to satisfy these 
conditions through the RGEs study. Details of the used parameters 
in the analysis are addressed in Appendix B and an example of the results of this
study is presented in the right panel of Fig.~1.
  
\begin{figure}[t]
\begin{center}
\includegraphics[width=7.5cm]{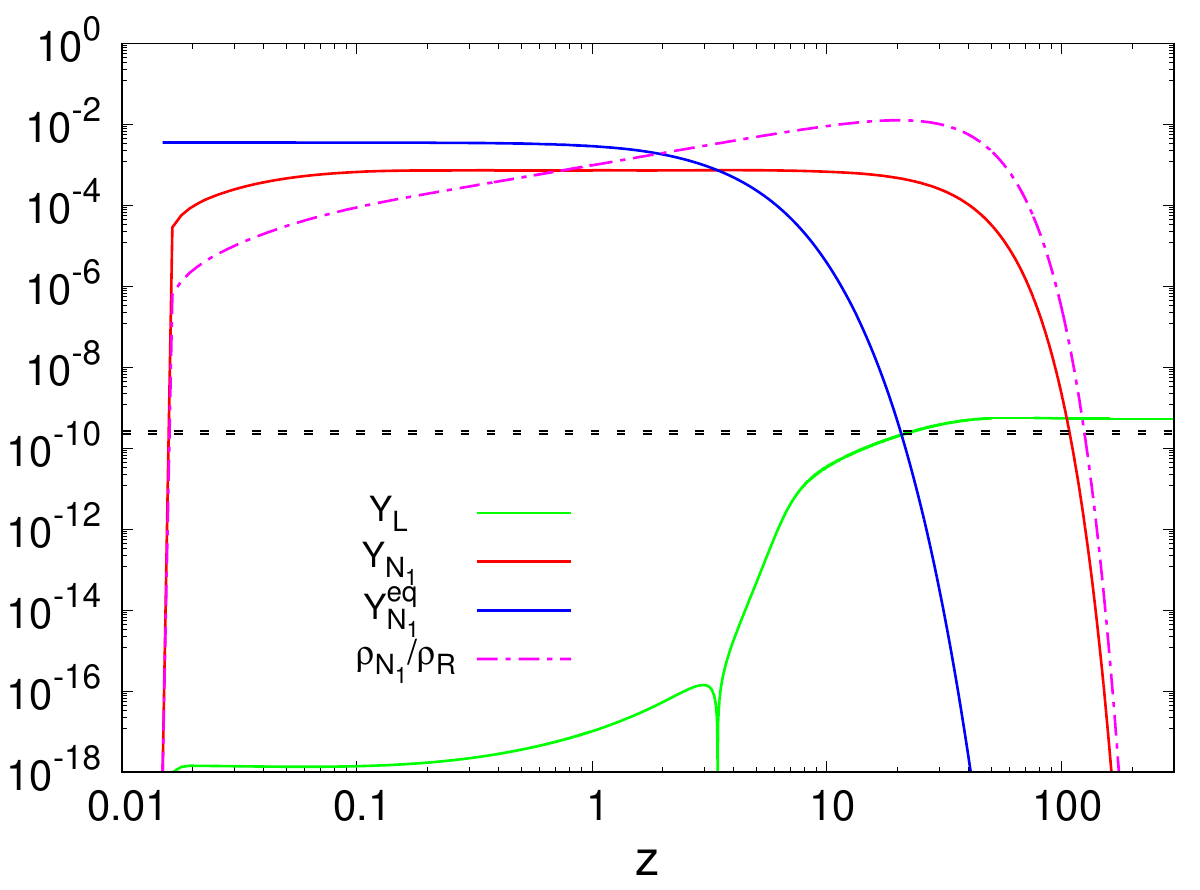}
\hspace*{3mm}
\includegraphics[width=7.5cm]{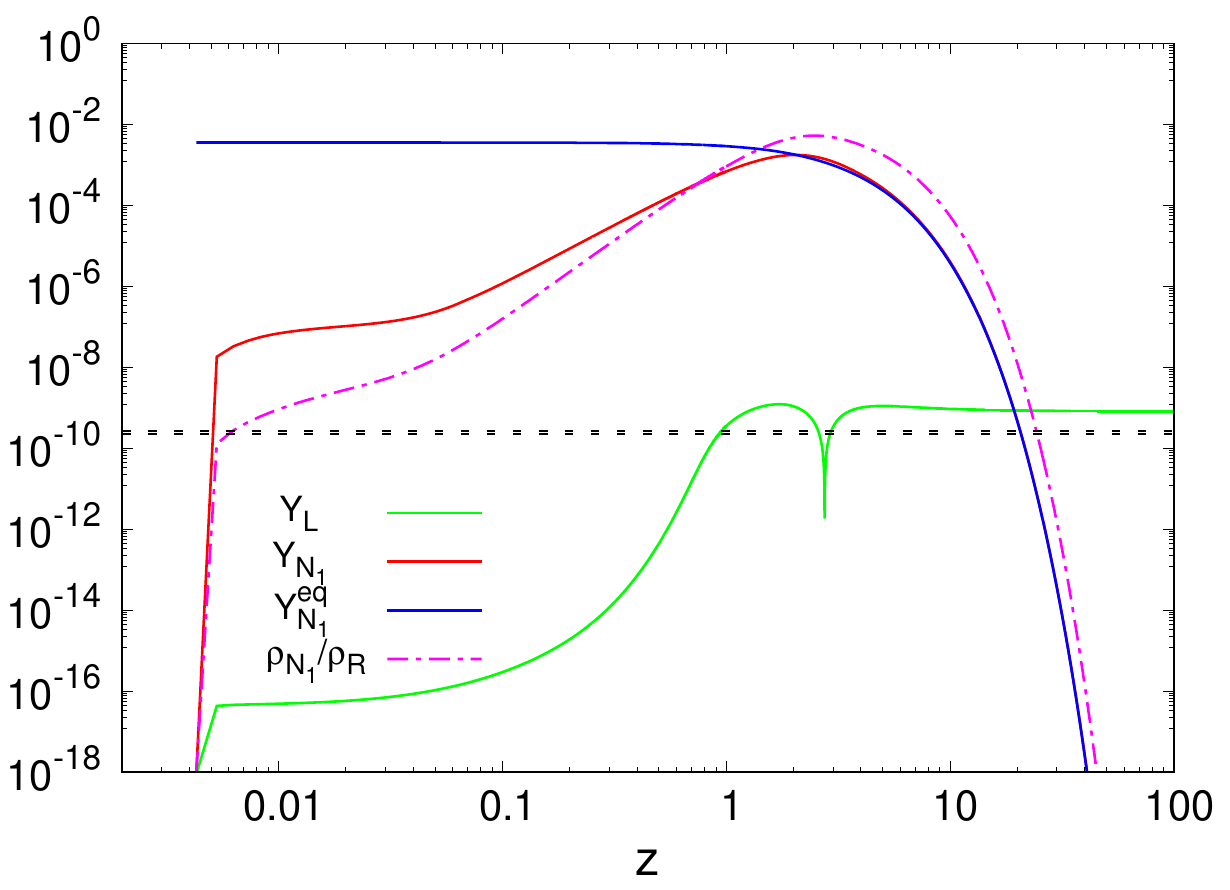}
\end{center}

{\footnotesize {\bf Fig.~5}~Evolution of $Y_{N_1}$ and $Y_L$ obtained as 
solutions of Boltzmann equations.  Results of the case (I) is shown 
in the left panel for $\kappa_{\phi\sigma}=10^{-4}$ and 
$\kappa_{\phi S}=\kappa_{\eta S}=0$.
Results of the the case (II) is shown in the right panel for 
$\kappa_{\phi S}=\kappa_{\eta S}=10^{-6}$ 
and $\kappa_{\phi\sigma}=\kappa_{\eta\sigma}=0$. 
Other parameters in each case are given in the text. 
Initial values for them are fixed as $Y_{N_1}=Y_L=0$ at $z=z_R$.
$\rho_{N_1}/\rho_{\rm R}$ represents a ratio of the energy density 
of $N_1$ to the one of radiation. }
\end{figure}

Solutions of the Boltzmann equations in  the cases (I) and (II) are 
are shown in Fig.~5.  The lightest right-handed neutrino mass in each case is 
$M_{N_1}=7\times 10^6$ and $10^8$~GeV.
In both cases, the sufficient baryon number asymmetry is found to be produced.
The figure for the case (I) shows clearly that the present scenario works well.
$Y_{N_1}$ reaches a value near $Y_{N_1}^{\rm eq}$ through the scattering 
of the extra fermions as expected. Substantial out-of-equilibrium decay occurs 
 at $z>10$ to generate the lepton number asymmetry.
The delay of the decay due to the small $h_1$ could make the washout of 
the lepton number asymmetry ineffective. 
On the other hand,  we cannot definitely find a signature of the scenario 
in the case (II), where the $N_1$ mass is near the bound for which the usual 
leptogenesis can generate the required baryon number asymmetry in the original 
scotogenic model \cite{ks}. The figure shows that additional contribution to the 
$N_1$ production starts at $z\simeq 0.1$.
It is considered to be brought about by the $N_1$ inverse decay since it is
expected to become effective
around $z\sim \left(\frac{6.3\times 10^{-5}}{h_1}\right) 
\left(\frac{M_{N_1}}{10^8~{\rm GeV}}\right)$.  
The figure shows that it plays a main role for the $N_1$ production. 
As a result, the leptogenesis in this case results in the ordinary one where the 
lower mass bound of $N_1$ is $O(10^8)$ GeV. 

These results show that the model with suitable parameters can 
generate a sufficient amount of baryon number asymmetry through 
leptogenesis even if the reheating temperature is lower than $10^9$~GeV
as long as $M_{N_1}<T_R$ GeV is satisfied.
In the present model, both the right-handed neutrino mass and 
the extra fermion mass
are determined by $M_{N_j}=y_{N_j}w$ and $\tilde M_F=\delta y_Fw$.
Since $w$ is fixed as the $PQ$ symmetry breaking scale and 
$\delta>1$ is imposed for the realization of the substantial $CP$ phases in the CKM 
and PMNS matrices, their mass cannot be arbitrarily smaller than $10^9$ GeV 
under the condition that the scattering of the extra fermions to the right-handed 
neutrinos is effective.
Because of this reason, low scale leptogenesis, which can be a distinguishable feature 
of the model, tends to be allowed only for the case where the PQ symmetry breaking
occurs at a neighborhood of its lower bound.
Even in that case, successful leptogenesis is expected to be realized only in the range 
 $M_{N_1}>4\times10^6\left(\frac{y_F}{10^{-1.2}}\right)^{-1/2}$ GeV for the case 
$T_R>10^8$ GeV since $T$ in eq.~(\ref{scat}) should satisfy $T>\tilde M_F, M_{N_1}$ 
for the sufficient $N_1$ production.

\begin{figure}
\begin{center}
\includegraphics[width=7.5cm]{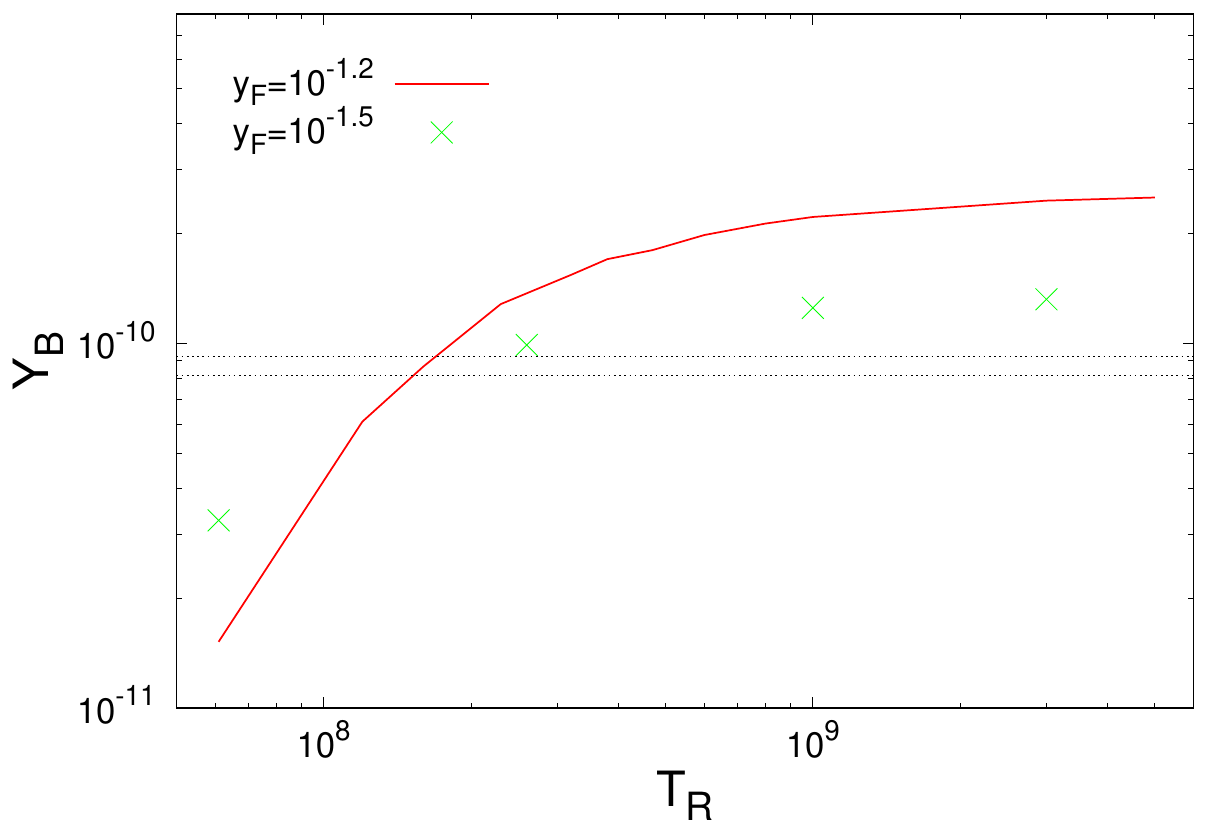}
\hspace*{5mm}
\includegraphics[width=7.5cm]{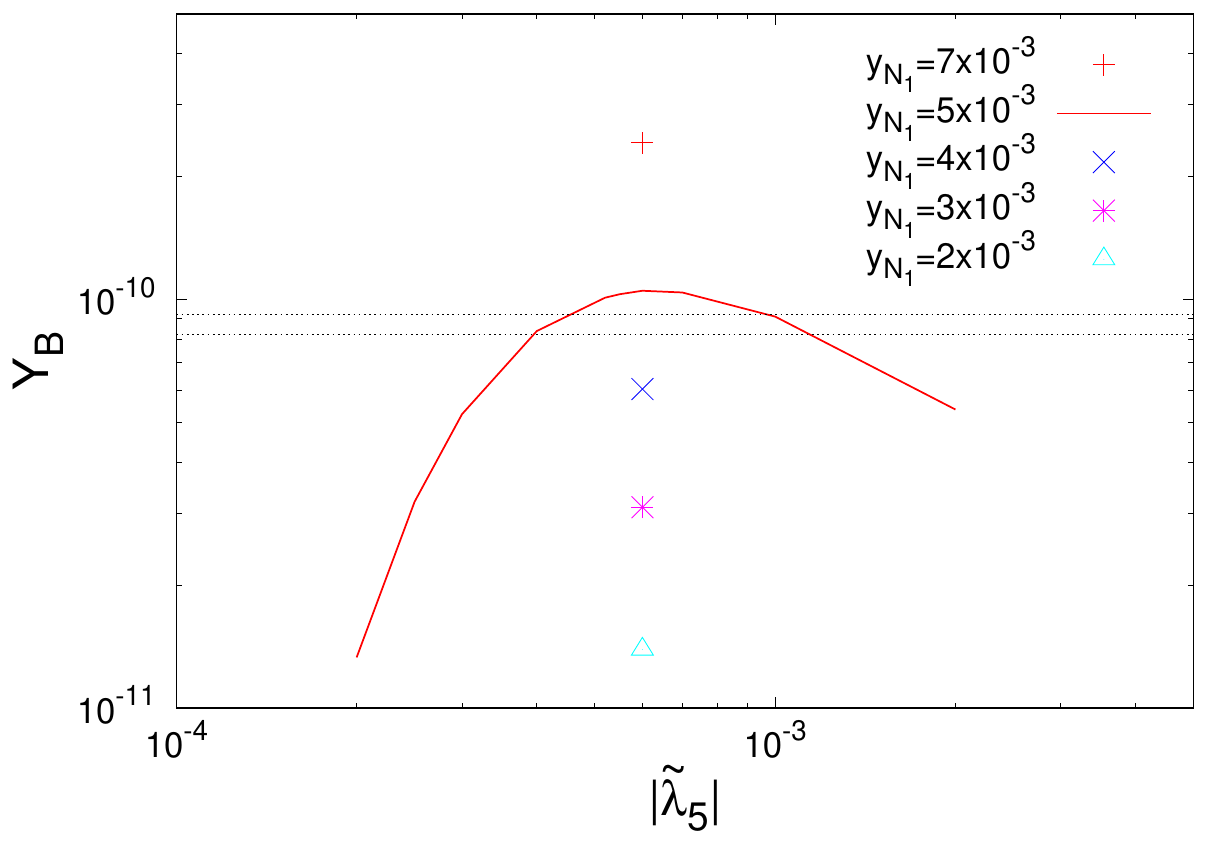}
\end{center}
\vspace*{-5mm}
\footnotesize{{\bf Fig.~6}~~Left : baryon number asymmetry generated 
at the expected reheating temperature $T_R$ (GeV) in the case (I). 
It is plotted by a red solid line for
$y_D=10^{-1.2}$ and by green crosses for $y_D=10^{-1.5}$. 
Right : baryon number asymmetry generated at $T_R=10^9$ GeV for various 
values of $\tilde\lambda_5$ and $y_{N_1}$ in the case (I). 
In both panels, other parameters are fixed at the ones given in (\ref{para}). }
\end{figure}

In Fig. 6, we show the baryon number asymmetry $Y_B$ generated  in the case (I) 
varying the values of relevant parameters. In the left panel, $Y_B$ is plotted 
as a function of the reheating temperature, which is fixed by the inflaton composition and 
its coupling with $\phi$ and $\eta$. Two values of $y_F$ are used in this plot.
The $N_1$ production in the present scenario depends on the reheating 
temperature and the couplings $y_F,~ y_{N_1}$.
A red solid line representing $Y_B$ is expected for 
the parameters given in eq.(\ref{para}) for the case (I).
It becomes larger and reaches an upper bound $Y_B\simeq 2.5\times 10^{-10}$ 
when the reheating temperature increases to $10^{10}$~GeV.   
This behavior can be understood if we take into account that
the equilibrium number density of extra fermions are suppressed by
the Boltzmann factor at lower reheating temperature and then 
the $N_1$ production due to the scattering of extra fermions is suppressed.
We also plot $Y_B$ for a smaller value of $y_F$ by green crosses at some typical
$T_R$. They show that $Y_B$ takes smaller values for a smaller $y_F$ 
since the $N_1$ production cross section is proportional 
to $y_F^2$.\footnote{Since the effect of Boltzmann suppression 
caused by its mass $\tilde M_F=\delta y_Fw$ could be dominant at lower $T_R$ 
compared with the effect on the cross section, the smaller $y_F$ gives a larger 
$Y_B$ at $T_R<10^8$ GeV in this case.} 
In the right panel,  $Y_B$ is plotted by varying
$|\tilde\lambda_5|$ and $y_{N_1}$.  A red solid line represents it as a function of 
$|\tilde\lambda_5|$ for a fixed $y_{N_1}=5\times 10^{-3}$.
Since the neutrino oscillation data have to 
be imposed on eq.~(\ref{oscilcon}), Yukawa couplings $h_{2,3}$ are settled 
by $|\tilde\lambda_5|$, $y_{N_2}$ and $y_{N_3}$.
The $CP$ asymmetry $\varepsilon$ and the washout of the generated 
lepton number asymmetry are mainly determined by $h_{2,3}$ for the fixed
$y_{N_2}$ and $y_{N_3}$ as found from eq.(\ref{epsilon}). 
Since a smaller $|\tilde\lambda_5|$ makes $h_{2,3}$ larger and then both $\varepsilon$ 
and washout larger, $Y_B$ takes a maximum value for a certain $|\tilde\lambda_5|$,
which is found in the figure.
We also plot $Y_B$ by varying $y_{N_1}$ for a fixed $|\tilde\lambda_5|$ in the same panel.
A smaller $y_{N_1}$ makes the $N_1$ production less effective for a fixed 
$y_F$ and then its lower bound is expected to appear for successful leptogenesis.
It gives the lower bound of $M_{N_1}$ as $\sim 4\times 10^6$ GeV
as predicted above.

Although other parameters are fixed at the ones given in (\ref{para}) 
in these figures, it is useful to give remarks on their dependence here.
If $\delta$ takes a larger value, the mass of extra fermions $\tilde M_F$ 
becomes larger to suppress the reaction density $\gamma_F$ 
due to the Boltzmann factor. 
As a result, the $N_1$ number density generated through 
the scattering becomes smaller and the resulting $Y_B$
also becomes smaller. If $h_1$ is much smaller, $N_1$ decay delays and 
the entropy produced through the decay of relic $N_1$ might dilute 
the generated lepton number asymmetry.

\subsection{Dark matter and isocurvature fluctuations}
This model has two DM candidates.
One  is the lightest neutral component of $\eta$ with $Z_2$ odd parity which is
an indispensable ingredient of the model. 
It is known to be a good DM candidate which does not cause any contradiction 
with known experimental data as long as its mass is in the TeV range 
where the coannihilation can be effective \cite{idm,highmass,ks}.
As found from Fig.~1, both the DM abundance and the DM direct 
search bound can be satisfied if the couplings $\tilde\lambda_3$ and $|\lambda_4|$ 
take suitable values of $O(1)$. 
Although these parameters could affect the perturbativity of the scalar quartic 
couplings through the radiative corrections, we can safely escape such problems
in certain parameter regions.
The results obtained for the case (I) in the previous part 
are derived by supposing that the required DM is $\eta_R$.

Axion is another promising candidate in the model. 
However, the axion could be a dominant component of DM only for 
$f_a\sim 10^{11}$~GeV although it depends on the contribution from 
the axion string decay \cite{axiondm}.
We consider the case (II) as such an example.
As described before, the PQ symmetry is spontaneously broken
during the inflation since the inflaton contains the radial component of $\sigma$.
As a result, the axion appears as the phase $\theta$ of $\sigma$.  
Since the axion potential is flat during the inflation, the axion gets 
a quantum fluctuation $\delta A=\left(H/2\pi\right)^2$ and it 
can cause isocurvature fluctuation in the CMB amplitude \cite{isocurv,isocurv2}.
A canonically normalized axion $A$ is defined by noting eq.~(\ref{canoninf}) as
\begin{equation}
\frac{\partial A}{\partial \theta}=\frac{\tilde\sigma}{\Omega^2}\sqrt{\Omega^2+
6\xi_\sigma\frac{\tilde\sigma^2}{M_{\rm pl}^2}}
\simeq\frac{\sqrt{\chi M_{\rm pl}}}{\tilde\xi_S^{1/4}}
\frac{|\kappa_{\sigma S}|}{\tilde\kappa_\sigma} \equiv\chi_{\rm iso}.
\end{equation} 
Since the axion interacts with other fields very weakly, it causes the isocurvature
fluctuation as the fluctuation of its number density $n_A$. 
The amplitude of its power spectrum can be expressed as
\begin{equation}
{\cal P}_i(k)=\Big\langle\left|\frac{\delta n_A}{n_A}\right|^2\Big\rangle
=\frac{H_k^2}{\pi^2\chi_{\rm iso}^2\langle \theta^2\rangle}.
\end{equation}  
Since the axion is only a source of the isocurvature fluctuation in this model, 
its fraction in the power spectrum is given as
\begin{equation}
\alpha=\frac{R_a^2{\cal P}_i(k)} {R_a^2{\cal P}_i(k)+{\cal P}_s(k)}
\simeq 8~\epsilon~\tilde\xi_S^{1/2}\frac{M_{\rm pl}}{\chi_k}
\frac{R_a^2}{\langle\theta^2\rangle}\left(\frac{\tilde\kappa_\sigma}{\kappa_{\sigma S}}\right)^2,
\end{equation}
where ${\cal P}_s(k)=A_s$ which is given in eq.(\ref{power}).
$R_a$ is a fraction of the axion energy density in the CDM and defined as 
$R_a=\Omega_a/\Omega_{\rm CDM}$. 
If we use a relation \cite{qcdaxion}
\begin{equation}
R_a=\frac{\langle\theta^2\rangle}{6\times 10^{-6}}\left(\frac{f_a}{10^{16}~{\rm GeV}}\right)^{7/6},
\end{equation}
we find 
\begin{equation}
\alpha=3.25\times 10^{-5}~\tilde\xi_S^{1/2}\frac{M_{\rm pl}}{\chi_k}
\left(\frac{55}{{\cal N}_k}\right)^2
\left(\frac{f_a}{10^{10}~{\rm GeV}}\right)^{7/6}R_a
\left(\frac{\tilde\kappa_\sigma}{\kappa_{\sigma S}}\right)^2.
\end{equation}
Since the Planck data constrain $\alpha$ as $\alpha\le 0.037$ at 
$k=0.05$~Mpc$^{-1}$ \cite{planck18}, 
we have a condition for the model to be consistent with the present observation of 
the isocurvature fluctuation in the CMB as 
\begin{equation}
R_a<\frac{67}{\tilde\xi_S^{1/2}}\frac{\chi_k}{M_{\rm pl}}\left(\frac{{\cal N}_k}{55}\right)^2
\left(\frac{10^{11}~{\rm GeV}}{w}\right)^{7/6}
\left(\frac{\kappa_{\sigma S}}{\tilde\kappa_\sigma}\right)^2,
\end{equation}
where $f_a=w$ is used.
In the case (I), this gives no constraint and the parameters used in the present study 
to estimate the reheating temperature in Fig.~4 and the baryon number asymmetry 
are consistent with the observational data. 
DM can be identified with the neutral component of $\eta$.
On the other hand, in the case (II), the isocurvature condition can be satisfied 
for $R_a<0.21$ if $|\kappa_{\sigma S}|/\tilde\kappa_\sigma=10^{-1.6}$ is assumed 
and for $R_a<0.034$ if $|\kappa_{\sigma S}|/\tilde\kappa_\sigma=10^{-2}$ is assumed.
The isocurvature constraint forbids the axion to be a dominant DM component and 
the neutral component of $\eta$ is required to play a role of the dominant DM 
in this case also. 

\section{Summary}
We have proposed a model which could give an explanation for 
the origin of the $CP$ phases in both the CKM and PMNS matrices
and the strong $CP$ problem.
It is a simple extension of the SM with vector-like extra fermions and several 
scalars. In order to control the couplings of new fields, global symmetry is imposed.  
If the $CP$ symmetry is spontaneously broken in a singlet scalar sector at
an intermediate scale, it can be transformed to the CKM and PMNS matrices 
through the mixing between the extra fermions and the ordinary quarks 
or the charged leptons. 
On the other hand, since the colored extra fermions play the same role 
as the ones in the KSVZ model for the strong $CP$ problem, 
the strong $CP$ problem could be solved through the PQ mechanism. 
After the symmetry breaking due to the singlet scalars, the leptonic sector of 
the model is reduced to the scotogenic model, which can explain the small 
neutrino masses and the DM abundance due to the remnant discrete symmetry
of the imposed symmetry.
Singlet scalars introduced to explain the $CP$ issues can play a role of inflaton
if it has a nonminimal coupling with the Ricci scalar.
We suppose this coupling is of order one. In that case,
although it gives the similar prediction for the scalar spectral index and 
the tensor to scalar ratio to the one of the Higgs inflation, 
reheating phenomena is different from it since the radiation domination starts 
just after the end of inflation.

The model has a notable phenomenological feature in addition to these.
The extra fermions which are introduced for the $CP$ issues could make the 
thermal leptogenesis generate the sufficient baryon number asymmetry 
even if the lightest right-handed neutrino mass is much lower than $10^9$~GeV, 
which is the well-known lower bound of the right-handed neutrino mass for 
successful leptogenesis in the ordinary seesaw scenario.  
Although the model allows low scale leptogenesis, it is difficult to
distinguish it from other thermal leptogenesis model experimentally.
However, if we consider its supersymmetric extension, it could give a possibility
to escape the gravitino problem. 
The model is constrained by the isocurvature fluctuation which is caused 
by the spontaneous breaking of the PQ symmetry during the inflation.
We find that its present observation can be consistent with the model even if 
the DM relic abundance is imposed on the model.  
Although the relic density of axion should be a small fraction of the DM, 
there is a neutral component of the inert doublet scalar as an alternative 
candidate of DM in the model.
It can explain the DM abundance just as in the scotogenic model without 
affecting other predictions of the model. 
It is remarkable that the model has potentiality to explain various issues 
in the SM although the model is rather simple.
It may deserve further study.

\section*{Appendix A~~ A simple example for $A_f$}
In this Appendix, we present a simple example which could bring 
about a phase in the CKM matrix.
In this example, we assume $w=10^9$ GeV and $u = 10^{11}$ GeV,
and also the relevant Yukawa couplings $h_d$, $y_d$ and
$\tilde y_d$ to be written by using real constant parameters as
\begin{equation}
h_d=c\left(\begin{array}{ccc}
\epsilon^4 &\epsilon^3 &p_1\epsilon^3\\
\epsilon^3 &\epsilon^2 & p_2\epsilon^2 \\
\epsilon^2 & p_3 & -p_3 \\
\end{array}\right), \quad
y_d=(0, y, 0), \quad \tilde y_d=(0, 0, \tilde y).
\label{yd}
\end{equation}
As long as $\epsilon$ satisfies $\epsilon < 1$, 
the down type quark mass matrix $m_d(\equiv h_d\langle\tilde\phi\rangle)$ 
has hierarchical mass eigenvalues. 
Here, we introduce $X_{ij}$ and $Y_{ij}$ whose definition is given as
\begin{eqnarray}
&&X_{ij}=1+p_ip_j 
+\frac{y^2+\tilde y^2p_ip_j
+y\tilde y(p_i+p_j)\cos 2\rho}{y^2+\tilde y^2}
\left(1-\frac{1}{\delta^2}\right), \nonumber \\
&&Y_{ij}=\frac{y\tilde y(p_i-p_j)\sin 2\rho}{y^2+\tilde y^2}
\left(1-\frac{1}{\delta^2}\right),
\end{eqnarray}
where $\delta$ is defined as $\delta=\tilde M_F/\mu_F$.
If we define $R_{ij}$ and $\theta_{ij}$ by using these quantities as
\begin{equation}
R_{ij}=\sqrt{X_{ij}^2+Y_{ij}^2}, \quad \tan\theta_{ij}=\frac{Y_{ij}}{X_{ij}},
\end{equation}
the component of eq.~(\ref{ckm}) is found to be expressed under the assumption
$\mu_D^2< {\cal F}_d{\cal F}_d^\dagger$ as
\begin{equation}
(A_d^{-1}m^2A_d)_{ij}=c^2\langle\tilde\phi\rangle^2 \epsilon_{ij}R_{ij}e^{i\theta_{ij}},
\label{diag}
\end{equation}
where $\epsilon_{ij}$ is defined as
\begin{equation}
\epsilon_{11}=\epsilon^6, \quad \epsilon_{22}=\epsilon^4, \quad
\epsilon_{33}=1, \quad \epsilon_{12}=\epsilon_{21}=\epsilon^5, \quad
\epsilon_{13}=\epsilon_{31}=\epsilon^3, \quad
\epsilon_{23}=\epsilon_{32}=\epsilon^2.
\end{equation}

By solving eq.~(\ref{diag}), we find that $A_d$ is approximately written as
\begin{equation}
A_d\simeq\left(
\begin{array}{ccc}
1 & -\lambda & \lambda^3\left(\frac{X_{23}}{|\alpha|^2X_{33}}e^{i\vartheta}
-\frac{X_{13}}{|\alpha|^3X_{33}}\right) \\
\lambda & 1 & -\lambda^2 \frac{X_{23}}{|\alpha|^2X_{33}}e^{i\vartheta}\\
 \lambda^3\frac{X_{13}}{|\alpha|^3X_{33}}&
 \lambda^2\frac{X_{23}}{|\alpha|^2X_{33}}e^{-i\vartheta} & 1 \\
\end{array}
\right),
\end{equation}
where the constants $\lambda$, $\alpha$, and $\vartheta$ are defined by
\begin{equation}
\alpha=\frac{X_{12}X_{33}-X_{13}X_{23}
e^{-i(\theta_{23}+\theta_{12}-\theta_{13})}}
{X_{22}X_{33}-X_{23}^2}, \quad
\lambda=|\alpha|\epsilon,\quad
\vartheta=\arg(\alpha) +\theta_{23}+\theta_{12}-\theta_{13}. 
\end{equation}
This expression shows that $A_d$ could have a nontrivial phase which 
gives the origin of the CKM phase. If the diagonalization matrix $O^L$ for a 
mass matrix of the up type quarks takes an almost diagonal form,
the CKM matrix could be obtained as $V_{\rm CKM}\simeq A_d$. 
As an example, if we assume $\cos\rho=\frac{\pi}{4}$ and fix other 
parameters as
\begin{eqnarray}
&&y_F=10^{-1.5}, \quad \delta=\sqrt{3}, \quad y=4\times 10^{-4}, \quad 
 \tilde y=2\times 10^{-4},   \nonumber \\
&&p_1=1.1, \quad p_2=-0.9, \quad p_3=1,  \quad \epsilon=0.2, \quad c=0.014,
\end{eqnarray}
then we obtain $\lambda=0.22$ and the Jarlskog invariant \cite{jar} 
as $J(\equiv {\rm Im}[ A_{12}A_{13}^\ast A_{23} A_{22}^\ast])=-1.6\times 10^{-6}$.
The mass eigenvalues for the down type quarks are obtained as
\begin{eqnarray}
&&m_d=\left\{X_{11}-\frac{X_{13}^2}{X_{33}}+|\alpha|^2\left(X_{22}
-\frac{X_{23}^2}{X_{33}}-2\right)\right\}^{1/2}\epsilon^3c\langle\tilde\phi\rangle 
\simeq 3.3~{\rm MeV}, \nonumber \\
&&m_s=\left(X_{22}-\frac{X_{23}^2}{X_{33}}\right)^{1/2}\epsilon^2c\langle\tilde\phi\rangle
\simeq 138~{\rm MeV}, \nonumber\\ 
&&m_b=X_{33}^{1/2}c\langle\tilde\phi\rangle\simeq 4.2~{\rm GeV}.
\end{eqnarray} 
Although a diagonalization matrix $A_e$ for the charged lepton sector 
may be considered to take the same form as $A_d$, it is not favorable 
for a large $CP$ phase
in the PMNS matrix. In that case, since $A_e$ is the nearly diagonal, large flavor mixing 
has to be caused only by the neutrino sector and the Dirac $CP$ phase in the PMNS 
matrix becomes small as a result. A large $CP$ phase in the PMNS matrix requires
$A_e$ to have rather large off-diagonal elements also.

\section*{Appendix B~~ RGEs for coupling constants}
In order to examine both the stability and the perturbativity of the model 
from the weak scale to the Planck scale, we have to know the running of the 
coupling constants in eq.~(\ref{model}).
If we fix these coupling constants at an intermediate scale 
and solve the RGEs to the Planck scale, we can find their values 
throughout the scale. The one-loop RGEs of the relevant coupling constants 
are given as

\begin{eqnarray}
&&16\pi^2\mu\frac{\partial \lambda_1}{\partial\mu}=24\lambda_1^2
+\lambda_3^2+(\lambda_3+\lambda_4)^2 + \kappa_{\phi\sigma}^2
+ \kappa_{\phi S}^2
+\frac{3}{8}\left(3g^4+g^{\prime 4}+2g^2g^{\prime 2}\right) \nonumber \\
&&\hspace*{20mm}-3\lambda_1\left(3g^2+g^{\prime 2}-4h_t^2\right)-6h_t^4,  \nonumber\\
&&16\pi^2\mu\frac{\partial \lambda_2}{\partial\mu}=24\lambda_2^2+\lambda_3^2
+(\lambda_3+\lambda_4)^2 + \kappa_{\eta\sigma}^2
+ \kappa_{\eta S}^2 
+\frac{3}{8}\left(3g^4+g^{\prime 4}+2g^2g^{\prime 2}\right)
 \nonumber \\
&&\hspace*{20mm}-3\lambda_2\left(3g^2+g^{\prime 2}\right) +4\lambda_2\left(2h_2^2+3h_3^2\right)-8h_2^4-18h_3^4, \nonumber\\
&&16\pi^2\mu\frac{\partial \lambda_3}{\partial\mu}=2(\lambda_1+\lambda_2)
(6\lambda_3+2\lambda_4)
+4\lambda_3^2+2\lambda_4^2 +2\kappa_{\phi\sigma}\kappa_{\eta\sigma}+ 
 2\kappa_{\phi S}\kappa_{\eta S}\nonumber \\
&&\hspace*{20mm}+\frac{3}{4}\left(3g^4+g^{\prime 4}-2g^2g^{\prime 2}\right)
-3\lambda_3\left(3g^2+g^{\prime 2}\right)
+2\lambda_3\left(3h_t^2+2h_2^2+3h_3^2\right), \nonumber \\
&&16\pi^2\mu\frac{\partial \lambda_4}{\partial\mu}=4(\lambda_1+\lambda_2)\lambda_4+8\lambda_3\lambda_4+4\lambda_4^2
+3g^2g^{\prime 2}-3\lambda_4\left(3g^2+g^{\prime 2}\right) \nonumber \\
&&\hspace*{20mm}+2\lambda_4\left(3h_t^2+2h_2^2+3h_3^2\right),  \nonumber \\
&&16\pi^2\mu\frac{\partial \kappa_S}{\partial\mu}=20\kappa_S^2
+\kappa_{\sigma S}^2+2(\kappa_{\phi S}^2+\kappa_{\eta S}^2)
+4\kappa_S\left[3(y_{d}^2+\tilde y_{d}^2)+y_{e}^2+\tilde y_{e}^2\right] \nonumber \\
&&\hspace*{20mm}-2\left[3(y_{d}^2+\tilde y_{d}^2)^2)
+(y_{e}^2+\tilde y_{e}^2)^2\right], \nonumber \\
&&16\pi^2\mu\frac{\partial \kappa_\sigma}{\partial\mu}=20\kappa_\sigma^2
+\kappa_{\sigma S}^2+2(\kappa_{\phi \sigma}^2+\kappa_{\eta \sigma}^2)
+4\kappa_\sigma\left(3y_D^2+y_E^2+\frac{1}{2}y_{N_3}^2\right)-
2\left(3y_D^4+y_E^4+\frac{1}{2}y_{N_3}^4\right), \nonumber \\
&&16\pi^2\mu\frac{\partial \kappa_{\sigma S}}{\partial\mu}= 4\kappa_{\sigma S}^2+
8(\kappa_S+\kappa_{\sigma})\kappa_{\sigma S}+
2(\kappa_{\phi S}\kappa_{\phi\sigma}
+\kappa_{\eta S}\kappa_{\eta\sigma})\nonumber \\
&&\hspace*{20mm}+2\kappa_{\sigma S}\left[3(y_{d}^2
+\tilde y_{d}^2)+y_{e}^2+\tilde y_{e}^2
+3y_D^2+y_E^2+\frac{1}{2}y_{N_3}^2\right] 
-4[3y_D^2(y_d^2+\tilde y_d^2)+y_E^2(y_e^2+\tilde y_e^2)], \nonumber \\
&&16\pi^2\mu\frac{\partial \kappa_{\phi\sigma}}{\partial\mu}=
4\kappa_{\phi\sigma}^2+
2\kappa_{\sigma S}\kappa_{\phi S}+2\kappa_{\eta\sigma}(2\lambda_3+\lambda_4)
+4\kappa_{\phi\sigma}(3\lambda_1+2\kappa_\sigma)\nonumber \\
&&\hspace*{20mm}+2\kappa_{\phi\sigma}\left(3y_D^2+y_E^2+\frac{1}{2}y_{N_3}^2+
3h_t^2-\frac{9}{4}g^2
-\frac{3}{4}g^{\prime 2}\right), \nonumber \\
&&16\pi^2\mu\frac{\partial \kappa_{\eta\sigma}}{\partial\mu}=4\kappa_{\eta\sigma}+
2\kappa_{\sigma S}\kappa_{\eta S}+2\kappa_{\phi\sigma}(2\lambda_3+\lambda_4)
+4\kappa_{\eta\sigma}(3\lambda_2+2\kappa_\sigma) \nonumber \\
&&\hspace*{20mm}+2\kappa_{\eta\sigma}\left(3y_D^2+y_E^2+\frac{1}{2}y_{N_3}^2 +2h_2^2+3h_3^2
 -\frac{9}{4}g^2-\frac{3}{4}g^{\prime 2}\right)-4(h_2^2+h_3^2)y_{N_3}^2 \nonumber \\
&&16\pi^2\mu\frac{\partial \kappa_{\phi S}}{\partial\mu}=4\kappa_{\phi S}^2+
2\kappa_{\sigma S}\kappa_{\phi \sigma}+2\kappa_{\eta S}(2\lambda_3+\lambda_4)
+4\kappa_{\phi S}(3\lambda_1+2\kappa_S) \nonumber \\
&&\hspace*{20mm}+2\kappa_{\phi S}\left[3(y_{d}^2+\tilde y_{d}^2)
+y_{e}^2+\tilde y_{e}^2+3h_t^2 -\frac{9}{4}g^2-\frac{3}{4}g^{\prime 2}\right], 
\nonumber 
\end{eqnarray}
\begin{eqnarray}
&&16\pi^2\mu\frac{\partial \kappa_{\eta S}}{\partial\mu}=4\kappa_{\eta S}^2+
2\kappa_{\sigma S}\kappa_{\eta\sigma}+2\kappa_{\phi S}(2\lambda_3+\lambda_4)+
4\kappa_{\eta S}(3\lambda_2+2\kappa_S)\nonumber\\
&&\hspace*{20mm}+2\kappa_{\eta S}\left[3(y_{d}^2+\tilde y_{d}^2)
+y_{e}^2+\tilde y_{e}^2+2h_2^2+3h_3^2
-\frac{9}{4}g^2-\frac{3}{4}g^{\prime 2}\right], \nonumber \\
&&16\pi^2\mu\frac{\partial y_{d}}{\partial\mu}=y_{d}\left[-8g_s^2
-\frac{2}{3}g^{\prime 2}+\frac{1}{2}y_D^2
+4y_{d}^2+3\tilde y_{d}^2+y_{e}^2+\tilde y_{e}^2\right], \nonumber \\\
&&16\pi^2\mu\frac{\partial\tilde y_{d}}{\partial\mu}=\tilde y_{d}
\left[-8g_s^2-\frac{2}{3}g^{\prime 2}+\frac{1}{2}y_D^2
+3y_{d}^2+4\tilde y_{d}^2+y_{e}^2+\tilde y_{e}^2\right], \nonumber \\
&&16\pi^2\mu\frac{\partial y_{e}}{\partial\mu}=y_{e}\left[-6g^{\prime 2}+\frac{1}{2}y_E^2
+3(y_{d}^2+\tilde y_{d}^2)+2y_{e}^2+\tilde y_{e}^2\right], \nonumber \\
&&16\pi^2\mu\frac{\partial \tilde y_{e}}{\partial\mu}=\tilde y_{e}\left[-6g^{\prime 2}+\frac{1}{2}y_E^2
+3(y_{d}^2+\tilde y_{d}^2)+y_{e}^2+2\tilde y_{e}^2\right], \nonumber \\
&&16\pi^2\mu\frac{\partial y_D}{\partial\mu}=y_D\left(-8g_s^2-\frac{2}{3}g^{\prime 2}
+4y_D^2+y_E^2+\frac{1}{2}y_d^2+\frac{1}{2}\tilde y_d^2+\frac{1}{2}y_{N_3}^2\right), \nonumber \\
&&16\pi^2\mu\frac{\partial y_E}{\partial\mu}=y_E\left(-6g^{\prime 2}
+3y_D^2+2y_E^2+\frac{1}{2}y_e^2+\frac{1}{2}\tilde y_e^2+\frac{1}{2}y_{N_3}^2\right), \nonumber \\
&&16\pi^2\mu\frac{\partial y_{N_3}}{\partial\mu}=y_{N_3}\left[3y_D^2+y_E^2
+\frac{3}{2}y_{N_3}^2+2(h_2^2+h_3^2)\right], \nonumber \\
&&16\pi^2\mu\frac{\partial h_2}{\partial\mu}=h_2\left(-\frac{9}{4}g^2-
\frac{3}{4}g^{\prime 2}+5h_2^2+3h_3^2\right),
 \nonumber\\
&&16\pi^2\mu\frac{\partial h_3}{\partial\mu}=h_3\left(-\frac{9}{4}g^2-
\frac{3}{4}g^{\prime 2}+2h_2^2+\frac{15}{2}h_3^2+\frac{1}{2}y_{N_3}^2\right),
 \nonumber\\
&&16\pi^2\mu\frac{\partial h_t}{\partial\mu}=h_t\left(\frac{9}{2}h_t^2-8g_s^2-\frac{9}{4}g^2
-\frac{17}{12}g^{\prime 2}\right),  \nonumber\\
&&16\pi^2\mu\frac{\partial g_s}{\partial\mu}=-\frac{19}{3}g_s^3, \qquad
16\pi^2\mu\frac{\partial g}{\partial\mu}=-3g^3, \qquad
16\pi^2\mu\frac{\partial g^\prime}{\partial\mu}=\frac{79}{9}g^{\prime 3},
\end{eqnarray}
where $g_s$, $g$, and $g^\prime$ are the gauge coupling constants of the SM.
In these equations, we assume eq.~(\ref{tribi}) for neutrino Yukawa couplings
and eq.~(\ref{yd}) for $y_f$ and $\tilde y_f$, and also only
$y_{N_3}$ is taken into account. Contributions from $V_b$ are also neglected
in these RGEs.  

Initial values of the coupling constants which are used in the RGE study of the right panel 
of Fig.~1 are taken to be the same ones which are used in the analysis of the leptogenesis 
for the case (I).  
A part of them are fixed at the weak scale as 
\begin{eqnarray}
&&\tilde\lambda_1=0.13, \quad \tilde\lambda_2=0.1,  \quad \tilde\lambda_3=0.445, \quad  
\lambda_4=-0.545, \quad \lambda_5=-10^{-3}, \nonumber \\
&&h_1= 6\times 10^{-7}, \quad h_2= 8.3\times 10^{-3} , \quad  h_3=3.9\times 10^{-3}, 
\end{eqnarray}
where $\tilde\lambda_{3}$ and $\lambda_4$ are fixed by the DM constraint shown 
in the left panel of Fig.~1 and 
$h_{2,3}$ are fixed by using the nautrino mass formula and the neutrino oscillation data.
Remaining ones are fixed at a scale $\bar M$ as 
\begin{eqnarray}
&&\tilde\kappa_S=10^{-8},   \quad \tilde\kappa_\sigma=10^{-4.5}, \quad  |\kappa_{\sigma S}|=10^{-6.1}, 
\quad \kappa_{\phi\sigma}=\kappa_{\eta\sigma}=10^{-4},  \quad 
\kappa_{\phi S}=\kappa_{\eta S}=10^{-7}\nonumber \\
&& y_{N_1}=7\times 10^{-3}. \quad y_{N_2}=2\times 10^{-2}, \quad y_{N_3}=4\times 10^{-2},  
\nonumber\\ 
&&y_D=y_E=10^{-1.2}, \quad y_e=y_d=8\times 10^{-4}, \quad \tilde y_e=\tilde y_d=4\times 10^{-4}.
\end{eqnarray}
These satisfy the imposed conditions (\ref{stab1}), (\ref{uw}), (\ref{stab2}), (\ref{ks}) and (\ref{kap}).

\section*{Acknowledgements}
This work is partially supported by a Grant-in-Aid for Scientific Research (C) 
from Japan Society for Promotion of Science (Grant No. 18K03644).
N.~S.~R is supported by \textit{Program 5000 Doktor} under Ministry of 
Religious Affairs (MORA), Republic of Indonesia.

\newpage
\bibliographystyle{unsrt}

\end{document}